%% file: main.tex
\documentclass[onecolumn,11pt]{article}
\usepackage[top=1in, bottom=1in, left=1in, right=1in]{geometry}

\usepackage{Harms_APSDFD_2022}
\usepackage{chapterfolder}

\graphicspath{ {../FIGS/} }

\title{{\fontsize{16}{16}\selectfont \textbf{Lagrangian Gradient Regression for the Detection of Coherent Structures from Sparse Trajectory Data}}} 

\author{\normalsize{Tanner D. Harms$^{1*}$, Steven L. Brunton$^{2}$, Beverley J. McKeon$^{1,3}$}\\
\footnotesize{$^1$ Graduate Aerospace Laboratories, California Institute of Technology, Pasadena, CA 91106, United States} \\
\footnotesize{$^2$ Department of Mechanical Engineering, University of Washington, Seattle, WA 98195, United States} \\
\footnotesize{$^3$ Stanford University, Center for Turbulence Research, Palo Alto, CA, United States\vspace{-.2in}}
}

\date{}

\begin{document}

\maketitle

\blfootnote{$^*$ Corresponding author (tharms@caltech.edu).}
\vspace{-.2in}
\begin{abstract}
    Lagrangian Coherent Structures (LCS) are flow features which are defined to objectively characterize complex fluid behavior over a finite time regardless of the orientation of the observer.  Fluidic applications of LCS include geophysical, aerodynamic, biological, and bio-inspired flows---among others---and can be generalized to broader classes of dynamical systems.  One of the prevailing paradigms for identifying LCS involves examining continuum-mechanical properties of the underlying flow.  Such methods, including finite-time Lyapunov exponent (FTLE) and Lagrangian-averaged vorticity deviation (LAVD) analyses, provide consistent and physically-meaningful results but require expensive computations on a dense array of numerically integrated trajectories.  Faster, more robust, sparse methods, on the other hand, are typically non-deterministic and require \textit{a-priori} intuition of flow field behavior for interpretation.  If LCS are to be used in the decision-making protocol of future autonomous technologies, a bridge between dense and sparse approaches must be made.  This work begins to address this goal through the development of Lagrangian gradient regression (LGR), which enables the computation of deformation gradients and velocity gradients from sparse data via regression.  Using LGR, velocity gradients, flow map Jacobians (thereby FTLE), and rotational Lagrangian metrics like the LAVD are accurately computed from sparse trajectory data which is orders of magnitude less dense than traditional approaches require.  Moreover, there is no need to compute velocity or numerically differentiate at any point when using LGR.  Therefore, it is a purely Lagrangian technique.
\end{abstract}

\input{0_Intro.tex}

\input{1_BackgroundTheory.tex}

\input{2_TimeDerivatives.tex}

\input{3_LocallyLinearRegressions.tex}

\input{4_EllipticLCS.tex}

\input{5_RandomParticleFields.tex}

\input{6_SummaryAndConclusions.tex}

\FloatBarrier
\clearpage
\input{A_Appendices.tex}

\bibliographystyle{unsrt}  
\bibliography{LLR2023_Refs_Updated}

\end{document}

%% file: 0_Intro.tex
\section{Introduction \label{sec:intro}}

Studying a dynamical system by observing the trajectories of tracers carried in the underlying flow is to consider it in the Lagrangian frame.  Many modern applications naturally admit such a description.  GPS outfitted ocean drifters~\cite{Lumpkin_SurfaceDrifterCoherentVortices, LumpkinCenturioni_DrifterData_2019, Putnam.Goni_TransportPredictionsPelagicSargassum_2020}, Arctic ice floes~\cite{Lopez-Acosta.Wilhelmus_IceFloeTracker_2019, Manucharyan.Wilhelmus_IceFloeMesoscale_2022}, and plastic pollution~\cite{vanSebille.Froyland_DynamicsOceanGarbageObservedDrifters_2012,vanSebille.Zika_LagrangianOceanAnalysis_2018,Hale.Zeng_GlobalPersectiveMicroplastics_2020}, for instance, are geophysical systems where objects are either directly studied or are used to study oceanic dynamics.  Particle flows are also common in the fluids laboratory where quantitative flow field measurements are taken using passive tracers with techniques like particle image velocimetry (PIV)~\cite{Raffel.Kompenhans_PIVbook_2018} or particle tracking velocimetry (PTV)~\cite{Nishino.Adrian_LPTV_1989, Schanz.Schroder_ShakeTheBox_2016, MalikBruecker_HighPrecisionPTV_1993, Maas.Willert_PTVVisualization_1993}.  Lagrangian methods are widely used even beyond the scope of fluid flows.  Some examples include the movement of cells in a developing embryo,~\cite{Mowlavi.Mahadevan_SparseNoisyLCS_2022}, network traffic~\cite{LiuZhang_TrafficFlowLAN_2016}, protein folding~\cite{Husic.Dabiri_sCSC_2019}, swarm dynamics~\cite{Chung.Kumar_SwarmSurvey_2018}, and numerical optimization~\cite{KennedyEberhart_PSOpaper_1995, Clerc_PSObook_2010}.  

Lagrangian particles represent observable instances of some underlying dynamical system that governs their behavior.  Coherent patterns exist in these dynamics, and identifying them is a critical task not only for scientific understanding, but also for engineering applications.  To this end much work has been devoted to extracting and characterizing coherent structures from high-dimensional data, for example in fluid mechanics~\cite{Taira.Yeh_ModalOverview_2017, Taira.Yeh_ModalApplicationsOutlook_2020}.  The theory of Lagrangian coherent structures uses the motion of embedded tracers to identify coherent structures~\cite{HallerYuan_LCSAndMixing_2000}.  Since the inception of the field, several excellent reviews and textbooks have been written, including Shadden (2011)~\cite{Shadden_LCS_2011}, Haller (2015)~\cite{Haller_LCS_2015}, Allshouse and Peacock (2015)~\cite{AllshousePeacock_LagrangianBasedMethods_2015},  Hadjighasem et al. (2017)~\cite{Hadjighasem.Haller_LCSCriticalComparison_2017}, and Haller (2023)~\cite{Haller_LCStextbook_2023}.  
LCS analysis also been widely applied in a variety of technical disciplines.  For example, LCS are often employed in the study of geophysical and atmospheric flows~\cite{SapsisHaller_app.geo_2009, Shadden.Marsden_app.geo_2009, Reniers.Olascoaga_app.geo_2010, Beron-Vera.Goni_app.geo_2008, Olascoaga.Kirwan_app.geo_2013, YuanHu_app.geo_2023, Filippi.Peacock_app.geo_2021, Nolan.Ross_app.geo_2019, Nolan.Powers_app.geo_2018}.  They have also been useful in characterizing structures in turbulent and unsteady flows of aerodynamic interest~\cite{Cao.Zhang_app.turbulence_2021, Pan.Zhang_app.turbulence_2009, Green.George_app.turbulence_2007, BauerKhinast_app.turbulence_2022, Sun.Bao_app.aero_2023, MullenersRaffel_app.aero_2012, Rockwood.Green_app.aero_2017, Wang.Deguchi_app.aero_2021, Ahmed.Hanifatu_app.aero_2023, Huang.Lin_app.aero_2022}, and in biomedical flows~\cite{Amahjour.Mancho_app.bio_2023, Yang.Wong_app.bio_2021, TewKai.Garcon_app.bio_2009, Nolan.Ross_app.bio_2020, Tallapragada.Schmale_app.bio_2011, ShaddenTaylor_app.bio_2008, Darwish.Kadem_app.bio_2021, PengDabiri_app.bio_2009, Shadden.Gerbeau_app.bio_2010}.  Moreover, since LCS theory applies to general dynamical systems, it has also been applied in studies beyond fluid dynamics~\cite{Lekien.Marsden_app.other_2007, Gawlik.Campagnola_app.other_2009, DiGiannatale.Bonfiglio_app.other_2018, Mowlavi.Mahadevan_SparseNoisyLCS_2022, Husic.Dabiri_sCSC_2019, LiuZhang_TrafficFlowLAN_2016}.

The properties of Lagrangian Coherent Structures make them particularly useful for the study of unsteady, transport-dominated flows.  One of the principal characteristics of an LCS is objectivity~\cite{Haller_LCS_2015, Gurtin.Anand_ContinuumMechanics_2010, TruesdellRajagopal_IntroToFluids_2000}, which guarantees that the observed quantity is invariant to translation and rotation.  It allows for reliable results on systems where the relationship between the observer and the flow is ambiguous or dynamic.  Another defining feature of LCS is the finite-time domain over which it is computed.  Due to their dependence on tracer trajectories, Lagrangian metrics must be computed over some time $t \in [t_0, t_0 + \Delta t],$ where $0<|\Delta t|<\infty$.  This distinguishes Lagrangian metrics from those computed from instantaneous flow snapshots such as the Okubo-Weiss criterion~\cite{Okubo_OkuboWeiss_1970,Weiss_OkuboWeiss_1991}, $Q$-criterion~\cite{Hunt.Moin_QCriteria_1988}, $\lambda_2$-criterion~\cite{JeongHussain_Lambda2_1995}, $\Gamma_1$- and $\Gamma_2$-criterion~\cite{Graftieaux.Grosjean_Gamma1Gamma2_2001}, and other field-based metrics~\cite{Chakraborty.Adrian_VortexIDReview_2005, Martins.Thompson_ObjectiveVortexCriteria_2016}, which do not give insight into the long-time dynamics of unsteady flows.  

Lagrangian coherent structures are typically computed using a dense grid of tracer particles to achieve reliable results~\cite{AllshousePeacock_LagrangianBasedMethods_2015}.  Such methods---referred to as dense approaches---are further categorized into geometric LCS~\cite{Haller_LCS_2015, Shadden.Marsden_FTLEProperties_2005} and probabilistic LCS~\cite{FroylandPadberg_ProbabilisiticAndGeometric_2009, Froyland.Monahan_TransportDynamicalSystems_2010, FroylandPadberg-Gehle_FTE_2012}.  The geometric approach uses tools from continuum mechanics and dynamical systems theory to ascribe physically meaningful values to material regions in the flow.  For instance, the finite-time Lyapunov exponent (FTLE) defines material surfaces of maximal material stretching over the analyzed time domain~\cite{Shadden.Marsden_FTLEProperties_2005}.  The probabilistic approach, on the other hand, bases its analysis on the Perron-Frobenius (or transfer) operator~\cite{FroylandPadberg-Gehle_FiniteTimeCoherentSets_2014}, which defines structures as regions where fluid material persists with high probability.  Both of these approaches have the advantage of producing identical results each time the analysis is performed on the same flow.  Geometric LCS have the additional advantage of interpretability, in the sense that structures have dynamical significance and do not require hyper-parameter tuning for identification.  

Despite the advantages of dense techniques, using LCS in scientific and decision-making applications remains challenging due to high computational cost and the difficulty of observing tracers over a meaningful duration.  Both geometric and probabilistic techniques require particle density that is greater than what is often achievable in the field or laboratory.  Therefore, numerical integration of artificial particles is necessary to achieve refined results.  Even if enough tracers can be observed, it is rarely possible in practical flows that the entire domain can be considered.  Particles often enter or leave the domain at the boundaries and often disappear for other reasons (for example, if the battery of an ocean drifter dies).  When this happens, the analysis is limited to the duration of the shortest particle trajectory.  

Sparse methods for computing LCS, which typically depend on data-driven clustering, have been developed in response to the first of these problems.  Many variations of sparse clustering methods have been developed for LCS identification, including the use of spectral clustering~\cite{Hadjighasem.Haller_SpectralClustering_2016}, fuzzy C-means~\cite{FroylandPadberg-Gehle_ClusterCmeans_2015}, graph coloring~\cite{Schlueter-KuckDabiri_CoherentStructureColoring_2017, Husic.Dabiri_sCSC_2019}, and DBSCAN~\cite{Mowlavi.Mahadevan_SparseNoisyLCS_2022}.  While these methods are able to achieve results with orders of magnitude fewer particles, they lack the determinism of dense methods and the interpretability of the geometric approach.  Users of sparse methods must select as a hyperparameter the number of structures to identify and must therefore have \textit{a-priori} information of the flow.  Additionally, the structures are not dynamically meaningful, and therefore require an informed operator to identify important flow features from the results.  So, while sparse LCS may be computed quickly, they are still not suitable for use in autonomous decision-making technologies due to user dependence.  Therefore, a principal direction of LCS research is to identify fast methods on sparse and noisy data that relate to physical properties of the flow.

Indeed, significant progress has already been made towards improving algorithmic performance on practical data sets.  To overcome the dependency on a structured array of particles, Lekien and Ross (2010)~\cite{LekienRoss_FTLEUnstructuredMesh_2010} implemented least-squares regression to compute the Jacobian on an unstructured mesh of particles.  Brunton and Rowley (2010)~\cite{BruntonRowley_FastComputationFTLE_2010} developed an algorithm utilizing flow map composition to speed up the calculations of FTLE fields, which decreased FTLE computation time for successive frames by reusing prior computations.  Raben et al. (2014)~\cite{Raben.Vlachos_FTLEonTRPIV_2014} built upon on these advances to develop strategies for computing FTLE directly from experimental particle trajectories, circumventing velocity field computations and improving computational efficiency. Recently, Mowlavi et al. (2022)~\cite{Mowlavi.Mahadevan_SparseNoisyLCS_2022} proposed a noise-robust method which augmented the regression approach of Lekien and Ross (2010)~\cite{LekienRoss_FTLEUnstructuredMesh_2010} by increasing the number of particle connections within the regression neighborhood and by adding Tikhonov regularization. Another approach to the sparse identification of physics-based LCS exists in the recent work by Haller et al. (2021-2023) where the trajectory stretching exponent (TSE) and trajectory rotation angle (TRA) have been developed as quasi-objective metrics that approximate traditional metrics like FTLE and polar rotation angle (PRA) and can be computed from single particle trajectories~\cite{Haller.Encinas-Bartos_Quasi-ObjectiveDiagnostics_2021, Encinos-Bartos.Haller_objectiveEddyViz_2022}.  The theoretical development of the deformation velocity by Kaszas et al. (2022)~\cite{Kaszas.Haller_objectiveCompVelocityField_2023} has enabled the TRE and TRA to be computed objectively when using velocity information from other observed particles in the flow domain~\cite{Aksamit.Rival_SparseStretchAndRot_2023}.  These methods have been seen to be quite effective at visualizing LCS on sparse data sets, but they remain an approximation of the original geometric LCS quantities.  Because metrics such as the FTLE and LAVD are closely linked to the evolution of the velocity gradient along a trajectory a significant barrier to their exact computation from sparse data is the lack of a purely Lagrangian approach to computing velocity gradients.  Without such a tool, dense velocity fields for each snapshot must be computed before velocity gradients along trajectories can be identified. Therefore, along with the persistent need for faster and more robust methods to compute FTLE fields, there exists a need for methods to compute elliptic LCS metrics directly from sparse data. 

This work proposes a method for computing hyperbolic, parabolic, and elliptic structures directly from sparse tracer trajectories.  While common geometric LCS analyses assume infinitesimal distances between particles, the proposed approach leverages flow map composition over consecutive segments of infinitesimal duration to enable accurate computation of typical LCS metrics from sparse data.  Additionally, it exploits the relationship of the short-time flow map Jacobian to the velocity gradient to provide a purely Lagrangian (i.e., directly from trajectory data) approach to compute velocity gradients and elliptic LCS metrics.  The specific contributions of this work include:  1) differentiation- and interpolation-free approximation of the velocity gradient tensor from particle trajectories alone; 2) accurate approximation of the flow map Jacobian over an arbitrary duration using sparse trajectories; 3) objective single-trajectory evaluation of the flow map Jacobian from flow data with reliable velocity gradients; and 4) computation of elliptic flow measures such as the LAVD from sparse data without velocity field information.

The structure of the paper is as follows: section \ref{sec:geomLCS} provides the prerequisite background and theory of geometric LCS and addresses common computational methods for computing geometric quantities.  Section \ref{sec:timederivs} considers the computation of geometric LCS quantities as the observation duration approaches zero.  This will lay the groundwork for the finite-time methods which come in sections \ref{sec:hyper_para} and \ref{sec:ellipLCS}, where methods for computing hyperbolic (stretch-based) and elliptic (rotation based) structures will be developed for sparse trajectories.  Finally, section \ref{sec:randdata} demonstrates the methods on random fields of particles to gauge the usefulness of the developed tools for experimental and field data.

%% file: 1_BackgroundTheory.tex
\section{Preliminary Theory of Geometric LCS \label{sec:geomLCS}}

Geometric LCS theory is an approach to characterize coherent structures according to the continuum mechanical properties of a flow \cite{Haller_LCS_2015}.  It is, perhaps, the most widely studied LCS subcategory, as it includes the study of finite-time Lyapunov exponents \cite{Haller_DistinguishedMaterialSurfaces_2001, Shadden.Marsden_FTLEProperties_2005} within its scope.  In this section, a brief background of geometric LCS theory is presented, followed by definitions of quantities required for extending the theory to sparse data.  The section is concluded by an example demonstrating the influence of sparsity on geometric LCS analysis.  

\begin{figure}[t!]
    \centering
    \includegraphics[width = 1\textwidth]{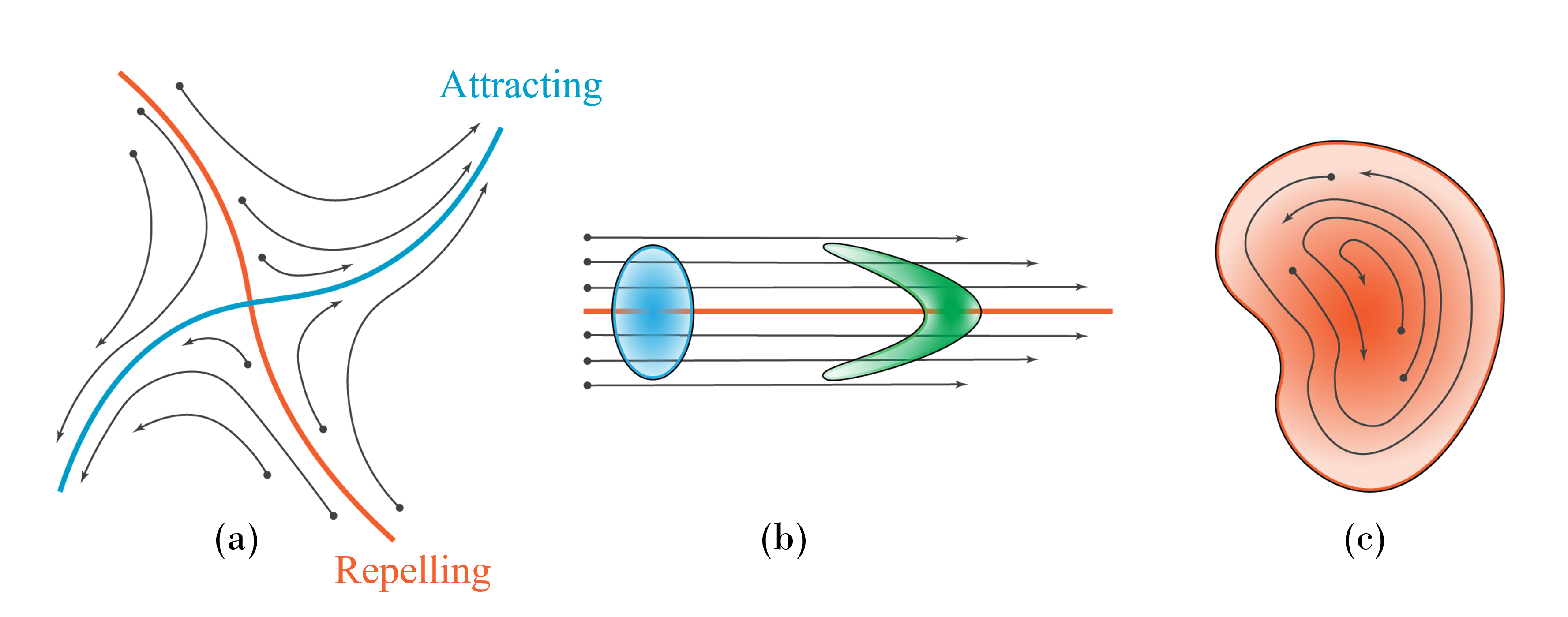}
    \caption{Categories of Geometric LCS.  (a) Hyperbolic LCS identify manifolds of attraction and repulsion. (b) Parabolic LCS are shear-driven and can be thought of as Lagrangian jet cores.  (c) Elliptic LCS consist of regions of coherent rotation.  }
    \label{fig:TypesOfLCS}
\end{figure}

Geometric LCS theory is thoroughly discussed in the seminal review and textbook by Haller~\cite{Haller_LCS_2015,Haller_LCStextbook_2023}. LCS are defined as codimension-1 manifolds (a line in 2-D or a surface in 3-D) that segment the flow into regions of quantitatively similar behavior over a finite observation period.  All geometric LCS can be classified into one of three categories: hyperbolic structures (defined as either attracting or repelling manifolds), parabolic structures (shear driven manifolds of maximal stretching often thought of as Lagrangian jet cores), and elliptic structures (families of coherent material loops existing between an outermost ring (or tube) and a central singularity \cite{HallerBeron-Vera_GeodesicTheory2D_2012, HallerBeron-Vera_CoherentVortices_2013}).  The three categories of structures can be visualized in figure \ref{fig:TypesOfLCS}.   

Elliptic LCS are often thought of as Lagrangian vortices defined over the time domain of interest.  Therefore, in this work, elliptic LCS are defined, not as codimension-1 surfaces, but rather as material volumes containing everything inside the outermost ring that is defined by the classical theory.  The conceptualization of these structures in the literature is varied, as some authors choose to lump parabolic structures into the elliptic category \cite{Mowlavi.Mahadevan_SparseNoisyLCS_2022}.  In this sense, elliptic LCS are defined as regions of flow where neighboring particles exhibit similar behavior (rotational, in the sense of a Lagrangian vortex, or translational as in the case of a Lagrangian jet core) over the time span.  In this work, however, parabolic and elliptic LCS will remain distinct. 

\subsection{The Flow Map}
The fundamental quantity in any geometric LCS analysis is the flow map, which is derived from the evolution of material elements within a flow.  The position $\mb{x}(t)$ of any passive tracer in a dynamical system is governed by the velocity field $\mb{v}(\mb{x}(t),t)$, where individual trajectories are solutions of 
\begin{equation}
    \dfrac{d}{dt}{\mb{x}(t)} = \mb{v}(\mb{x}(t),t).
    \label{eq:velocity}
\end{equation}
Let $D\subseteq\R^d$ be the domain of the system being studied, where $d=2,3$ in most cases of interest for fluid flows. 
It is often convenient to include the initial condition $\mb{x}_0$ and initial time $t_0$ as parameters in the particle trajectory $\mb{x}(t; t_0, \mb{x}_0)$.
The particle trajectory may be obtained by integrating through the velocity field according to
\begin{equation}
    \mb{x}(t; t_0, \mb{x}_0) = \mb{x}_0 + \int_{t_0}^t \mb{v}(\mb{x}(\tau; t_0, \mb{x}_0), \tau) d\tau,
\end{equation}
where  $\mb{x}(t_0; t_0, \mb{x}_0)=\mb{x}_0$ and where $\mb{x}(t; t_0, \mb{x}_0)$ constitutes a mapping
\begin{equation}
    \mb{F}_{t_0}^t : D\rightarrow D:\mb{x}_0 \mapsto \mb{F}_{t_0}^t(\mb{x}_0) = \mb{x}(t; t_0, \mb{x}_0).
    \label{eq:flowmap}
\end{equation}
For the sake of simplicity, dependence on initial conditions will be omitted from the notation unless it will aid in clarity.  Thus, $\mb{x}(t; t_0, \mb{x}_0)$ may be referred to as $\mb{x}(t)$, or even as $\mb{x}$, unless more specificity is required.

The mapping $\mb{F}_{t_0}^t(\mb{x}_0)$ is referred to as the flow map, and it brings material from an arbitrary reference location in $D$ to another spatial location in $D$ \cite{Gurtin.Anand_ContinuumMechanics_2010, TruesdellRajagopal_IntroToFluids_2000}.  The flow map is smooth in both space ($C^3$) and time ($C^1$) \cite{Shadden.Marsden_FTLEProperties_2005} and is a diffeomorphism with inverse
\begin{equation}
    \mb{x}_0 = (\mb{F}_{t_0}^t)^{-1}(\mb{x}(t)) = \mb{F}_{t}^{t_0}(\mb{x}(t)), 
\end{equation}
ensuring that no two reference points may occupy the same spatial point at a given time.  Other useful properties of the flow map stem from the local existence and uniqueness of solutions for initial value problems \cite{Shadden.Marsden_FTLEProperties_2005, Khalil_NonlinearSystems_2002}
\begin{subequations}
\begin{align}
    \mb{F}_{t_0}^{t_0}(\mb{x}_0) &= \mb{x}_0 \\
    \mb{F}_{t_0}^{t+s}(\mb{x}_0) & = \mb{F}_s^{t+s} \circ \mb{F}_{t_0}^s(\mb{x}_0) = \mb{F}_t^{t+s} \circ \mb{F}_{t_0}^t(\mb{x}_0).
\label{eq:ivpprops}
\end{align}
\end{subequations}
Thus, the flow map over the full time domain $[t_0,\, t]$ can be viewed as the composition of many intermediate flow maps.

\subsection{The Flow Map Jacobian}
The flow map Jacobian represents the deformation gradient of the flow map along a trajectory $\mb{x}$ starting at material point $\mb{x}_0$ and time $t_0$ and mapping to a spatial position $\mb{x}(t)$ at time $t$, and is given by
\begin{equation}
    D\mb{F}_{t_0}^t(\mb{x}_0) = \nabla_{\mb{x}_0} \mb{F}_{t_0}^t(\mb{x}_0),\qquad DF_{ij}(t; t_0, \mb{x}_0) = \dfrac{\partial x_i(t; t_0, \mb{x}_0)}{\partial x_{0,j}},\, \forall i,j \in [1,...,d].  \label{eq:flowjac}
\end{equation}

The flow map Jacobian governs the deviation of tracers within an $\epsilon$-neighborhood of the particle $\mb{x}_0$.  Consider the perturbation $\mb{y_0} = \mb{x}_0 + \Delta \mb{x}_0$, for $\Delta \mb{x}_0$ very small.  The relative position of the deformed perturbation is defined by
\begin{equation}
    \Delta \mb{x} = \mb{F}_{t_0}^t (\mb{y}_0) - \mb{F}_{t_0}^t (\mb{x}_0) =D\mb{F}_{t_0}^t(\mb{x}_0) \Delta \mb{x}_0 + \mathcal{O}(\norm{\Delta \mb{x}_0}^2) = \F(\mb{x}_0) \Delta \mb{x}_0 ,\, \text{ as } \Delta \mb{x}_0 \rightarrow 0, \label{eq:pertequation}
\end{equation}
where the right hand expressions are the result of Taylor Series approximation and the norm is Euclidean.  

For notational convenience, the argument of the tensors will be dropped unless it is ambiguous to do so.  Hence, $D\mb{F}_{t_0}^t \equiv D\mb{F}_{t_0}^t(\mb{x}_0)$, and so on.  Throughout this work, other tensors will be introduced, and it will be assumed that the argument is $\mb{x}_0$ unless otherwise stated.  

\subsection{Finite-Time Lyapunov Exponents}
The right Cauchy-Green stress tensor is another fundamental element of the theory of geometric LCS, and is defined as the Gram matrix of the flow map Jacobian 
\begin{equation}
    \mb{C}_{t_0}^t = \left(D\mb{F}_{t_0}^t\right)^\top D\mb{F}_{t_0}^t,
\end{equation}
where $(\cdot)^\top$ represents the matrix transpose.  $\mb{C}_{t_0}^t$ is a Gramian matrix, and therefore has the useful properties of being symmetric and positive semi-definite.  Since the continuum assumption of fluids enforces one-to-one behavior of the flow map, it is guaranteed that $\det D\mb{F}_{t_0}^t > 0$, and therefore that $\mb{C}_{t_0}^t$ is, in fact, positive definite and has $d$ real eigenvalues.  

Physically, the right Cauchy-Green stress tensor represents the squared change in local distances due to the flow map $\mb{F}_{t_0}^t$.  Thus, the maximum magnitude of the perturbation at time $t$ is determined by the $L^2$ norm on $D\mb{F}_{t_0}^t$
\begin{equation}
    \max \norm{\Delta \mb{x}} = \norm{\F}_{2} \norm{\Delta \mb{x}_0} = \sqrt{\lambda_{\max}(\C)}\norm{\Delta \mb{x}_0},
\end{equation}
where $\lambda_{\max}$ represents the largest eigenvalue of $\C$.  Additionally, due to the properties of the $L^2$ operator norm \cite{GolubVanLoan_MatComp_2013}, $\norm{\F}_{2}$ may be computed as the largest singular value of $\F$ and represents the maximal gain induced by the tensor.  
The scalar value $\sqrt{\lambda_{\max}(\C)}$ may be viewed as an exponential, such that 
\begin{equation}
    \sqrt{\lambda_{\max}(\C)} = e^{\sigma_{t_0}^t(\mb{x}_0) |\Delta t|},
\end{equation}
where $t_0 + \Delta t = t$.  The exponent $\sigma_{t_0}^t$ is the finite-time Lyapunov exponent, which represents the exponential growth rate of a linear deformation over the observation time.  Because $\sigma_{t_0}^t$ assumes linearity, it can only be reliably evaluated for finite $\Delta t$ if $\Delta \mb{x}$ is infinitesimal, or for finite $\Delta \mb{x}$ if $\Delta t$ is infinitesimal.  Therefore, it is typically computed according to
\begin{equation}
    \sigma_{t_0}^t = \lim_{\Delta \mb{x}_0\to 0}\frac{1}{|\Delta t|} \ln{ \sqrt{\lambda_{\max}(\C)} }.
    \label{eq:ftle}
\end{equation}

Both $\C$ and $\sigma_{t_0}^t$ are objective measures, though $\F$ is not.  For an extended discussion of objectivity and examples of the relevant computations, refer to appendix \ref{ap:objectivity}.

\subsection{Numerical Approaches for Computing Jacobians}

\begin{figure}[t!]
    \centering
    \includegraphics[width=1\textwidth]{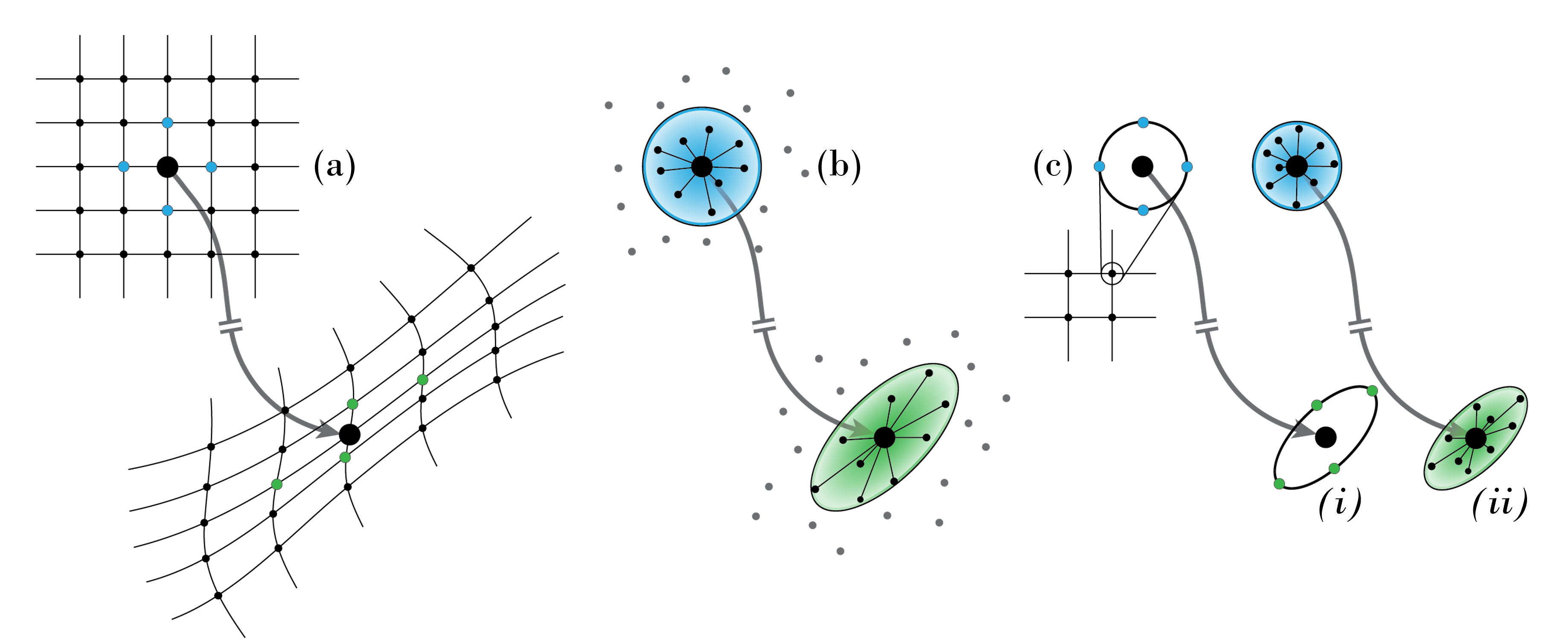}
    \caption{Diagrams of computation schemes for the flow map Jacobian in 2D.  (a) Finite-differences approach.  (b) Regression approach.  (c) Planet-satellite approach, which can be accomplished using either (i) finite-differences or (ii) regression. }
    \label{fig:CompSchemes}
\end{figure}

Geometric LCS are typically computed using either finite-differences or regression as the numerical engine.  A short description of these methods is provided here along with figure \ref{fig:CompSchemes}.  The finite differences approach to computing $\F$ was the first technique to be developed \cite{HallerYuan_LCSAndMixing_2000, Haller_DistinguishedMaterialSurfaces_2001, Shadden.Marsden_FTLEProperties_2005} and begins by artificially seeding massless particles throughout the flow domain in a fine mesh.  Velocity information for the flow---in the form of an analytical expression or snapshots from simulation or experiment---is then used to propagate the positions of the particles over some finite time domain $[t_0, t]$ to their deformed positions using a numerical integrator.  The gradient of the deformation (which is the flow map Jacobian at the initial time) is then approximated at each particle with respect to the initial positions using a finite-differencing scheme.  A schematic of this process is shown in figure \ref{fig:CompSchemes}(a).  

In some instances, the finite-differences algorithm for computing $\F$ is inefficient.  In circumstances where particle information is known on an unstructured mesh, for instance, superfluous steps are required to compute the LCS.  To address this, Lekien and Ross (2010) developed a method for computing $\F$ using Voronoi cell-weighted least-squares regression on unstructured meshes \cite{LekienRoss_FTLEUnstructuredMesh_2010}.   This method was adapted to experimental particle flows by Raben et al.\ in 2014, where regression was used on the observed particle trajectories themselves to compute LCS quantities.  The regression approach to LCS approximation is shown as a schematic in figure \ref{fig:CompSchemes}(b).

A third scheme was developed for this work with the intention of directly comparing various numerical approximation techniques.  This method, referred to as the planet-satellite approach and displayed in figure \ref{fig:CompSchemes}(c), involves specifying an array of test locations---perhaps on a uniform grid---then seeding particles around around them to locally perform computations either by finite-differences or by regression.  The resulting evaluation of $\F$ is then provided at the test locations on the initial grid. The primary advantage of this approach is that it allows multiple strategies to be evaluated on the grid of test points and compared directly.  

\subsection{Nonlinear Effects Dominate Sparse Data}
Computing accurate geometric LCS is predicated on achieving very small $\Delta \mb{x}_0$.  For a finite time domain, the validity of equation \ref{eq:pertequation} breaks down as $\Delta \mb{x}_0$ increases.  If an LCS analysis is still performed when $\Delta \mb{x}_0$ is large, the computed flow map Jacobian does not accurately describe the tracer deformation as the tracers it is computed from are non-negligibly influenced by flow nonlinearities.  

\begin{figure}[t!]
    \centering
    \includegraphics[width=1\textwidth]{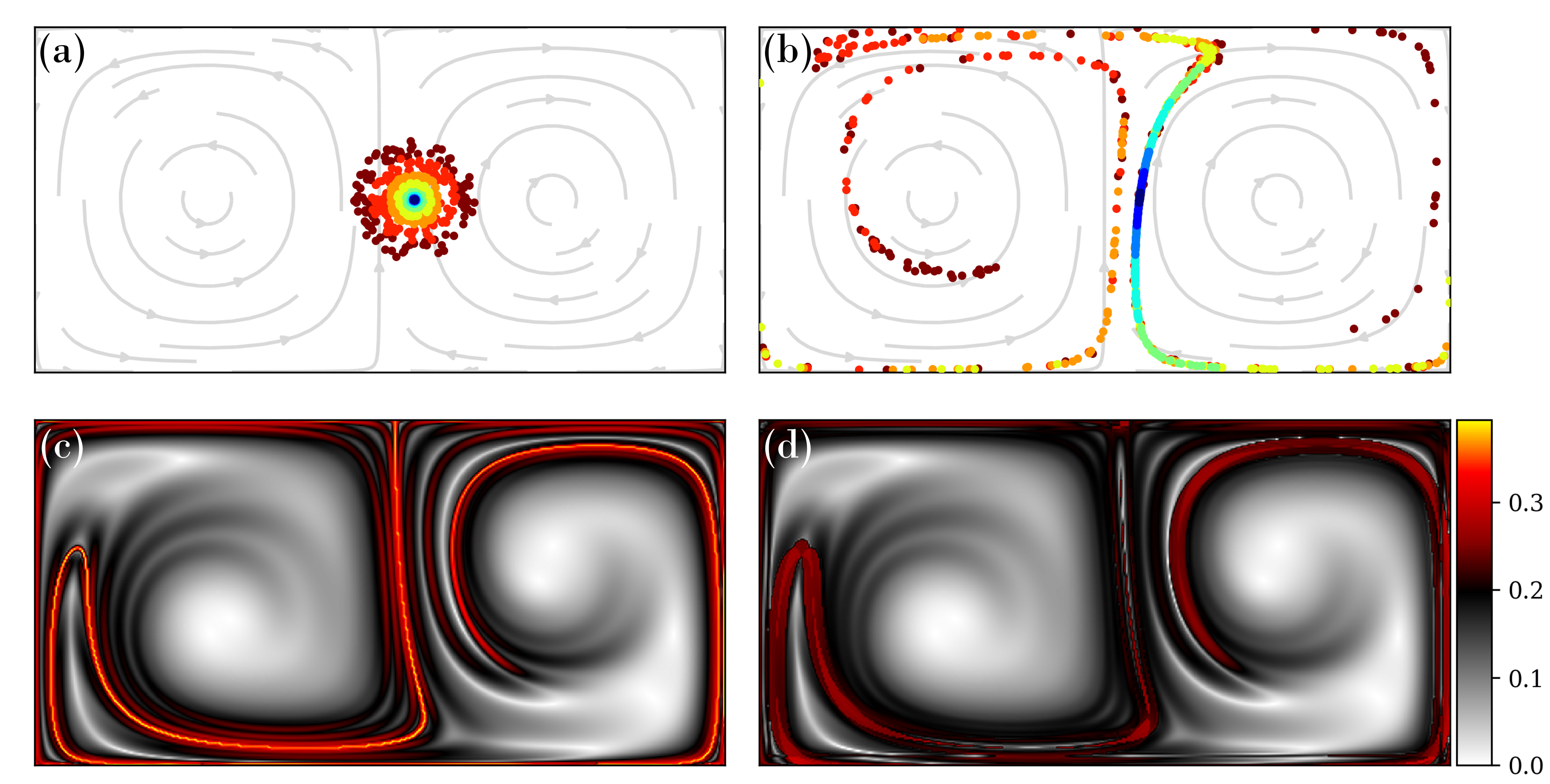}
    \caption{A demonstration of the nonlinear influence of the flow on accuracy of the Jacobian for particles with large initial radius.  (a) Particles organized by initial radius.  (b) Location of particles from (a) after 15 time units in the unsteady double gyre flow. (c) FTLE field evaluated at the level of grid spacing $\Delta \mb{p}_0$ ($\Delta \mb{p}_0 =\Delta \mb{x}_0 = 0.005$: light blue particles). (d) FTLE field evaluated with $\Delta x_0 = 5\Delta \mb{p}_0$ ($\Delta \mb{x}_0 = 0.025$: green particles).  All of the computations are performed using the planet-satellite approach with finite differences (figure \ref{fig:CompSchemes}(c.i))}
    \label{fig:LargeDx_Demo}
\end{figure}

To explore this principle, a case study is presented.  Consider the analytical flowfield of the unsteady double-gyre \cite{solomonGolub_Chaotic_1988,solomonGolub_Passive_1988}.  The velocity field is defined by
\begin{subequations}
\begin{align}
    u &= -\pi A \sin\left(\pi f(x,t)\right)\cos(\pi y),\\  
    v &= \pi A \cos\left(\pi f(x,t)\right)\sin(\pi y) \dfrac{\partial f(x,t)}{\partial x},
    \label{eq:doublegyre}
\end{align}
\end{subequations}
where
\begin{align*}
    f(x,t) &= a(t) x^2 + b(t)x, &   a(t) &= \epsilon \sin(\omega t),  & b(t) &= 1-2\epsilon \sin(\omega t),
\end{align*}
with parameters $\epsilon = 0.1$, $A = 0.1$, and $\omega = 2\pi/10$. The computational domain over which trajectories are integrated is $[0,\,2]$ in $x$ by $[0,\,1]$ in $y$ with $t_0 = 0$ and $t=15$, such that particles are advected through $1.5$ periods of the flow.  

To demonstrate the influence of $\Delta \mb{x}_0$ on approximating $\F$, tracer neighborhoods with varying radii are used to compute FTLE fields on the flow.  Figure \ref{fig:LargeDx_Demo}(a) shows a cloud of particles at time $t_0$ centered at $\mb{x}_0$ with varying radii indicated by color.  The particles are advected to the final positions (which are displayed in figure \ref{fig:LargeDx_Demo}(b)), where it is clear that the deformation of particles with large initial radius (green to red particles) cannot be reasonably approximated by a linear transformation.  However, as the initial radius decreases (blue particles), the deformation of the particles approximates an ellipsoid and can therefore be sufficiently described by a linear operator.  It is important to note that, as the integration time increases, the radius that can be accurately approximated by the flow map Jacobian decreases.

It is not clear simply by observing particle deformations how significantly the initial radius influences quantities used in identifying LCS.  Figures \ref{fig:LargeDx_Demo}(c) and (d) display the FTLE field computed on the flow using different radii in the $\F$ computations.  In both cases, the planet-satellite method is used with finite differences (figure \ref{fig:CompSchemes}(c.i)), where the set of evaluation locations $\{\mb{p}\}$ is a uniform grid and is kept the same between computations.  In figure \ref{fig:LargeDx_Demo}(c), the FTLE computations are performed using satellite spacing $\Delta \mb{x}_0 = 0.005$, which is equivalent to the grid spacing $\Delta \mb{p}$ and the radius of the light blue particles in (a) and (b).  Here, the FTLE field displays the sharp ridges that are consistent with the literature (see, for example, \cite{AllshousePeacock_LagrangianBasedMethods_2015}).  The computations in figure \ref{fig:LargeDx_Demo}(d) use $\Delta \mb{x}_0 = 5\Delta \mb{p} = 0.025$, which is the outer radius of the green particles in (a) and (b), and the graphical representation uses the same color mapping as for (c).  By visual inspection, it is seen that the regions from (c) with small FTLE values are largely unchanged, though the regions with large values (ie. the FTLE ridges) are muted and distorted. Indeed, the dominant ridge that appears in the center of the domain in (c) is difficult to identify in (d) as a result of the nonlinear warping effects of the flow.  As one of the prevailing definitions of hyperbolic LCS is the ridges of the FTLE field \cite{Shadden.Marsden_FTLEProperties_2005}, the field in (d) would identify spurious structures.  Additionally, as $\Delta \mb{x}_0$ and $\Delta t$ increase, the quality of the FTLE field decreases.

%% file: 2_TimeDerivatives.tex
\section{Lagrangian Gradient Regression \label{sec:timederivs}}

\begin{figure}[t!]
    \centering
    \includegraphics[width=1\textwidth]{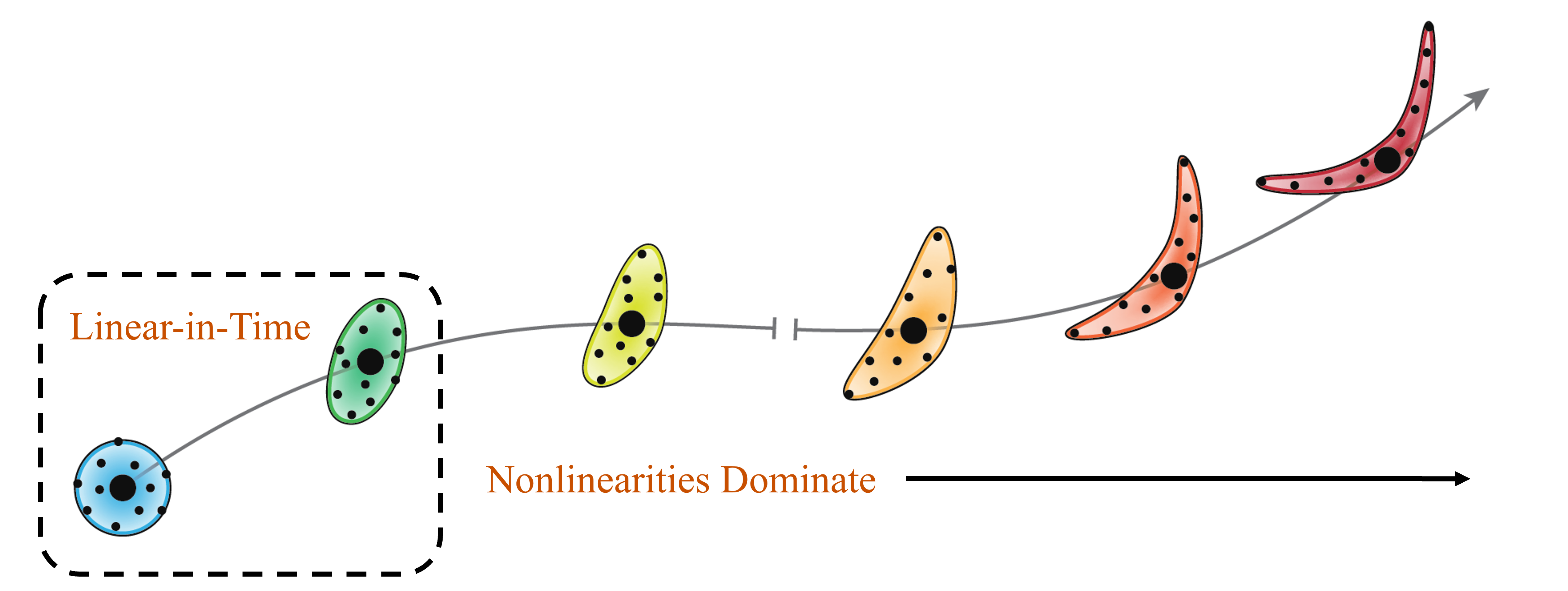}
    \caption{The deformation of a finite material neighborhood can always be approximated by a linear operator when using a sufficiently short time domain.  Colored ellipses represent the deformation of the blue circle along a trajectory indicated by the gray arrow.  Black dots represent observable tracers in the flow.}
    \label{fig:LIT_schematic}
\end{figure}

In the previous example it was demonstrated that the nonlinear deformations of material within a relatively large neighborhood will obscure the structures identified in LCS procedures.  However, if the deformations are considered over a sufficiently short duration, they can still be approximated by the flow map Jacobian as a result of continuity.  To visualize this, consider figure \ref{fig:LIT_schematic}, which represents the deformation of a finite neighborhood of material (on the left in a blue circle) over increasing durations.  In the short time ($\Delta t/ \mathcal{T} << 1$, where $\mathcal{T}$ is the fastest time-scale of the flow) the deformation of the material is approximately linear, transforming the circle into an ellipse.  However, as the system evolves, the deformations become increasingly complex (yellow, orange, and red blobs) and cannot be simply approximated by the flow map Jacobian because the higher-order terms from equation \ref{eq:pertequation} dominate.  

In this section, the characteristics of short-time material deformations are considered and are presented as the foundation of extensions to computing hyperbolic and elliptic structures over arbitrary durations, which is discussed in subsequent sections.  

\subsection{Evolution Equation of the Flow Map Jacobian}
The flow map Jacobian evolves according to a linear process defined by 
\begin{align}
    \dfrac{d}{d t} \F &= \nabla \mb{v}(\mb{x}(t),t) \F,
    \label{eq:linearProcessJacobian}
\end{align}
where $D\mb{F}_{t_0}^{t_0} = \mb{I}_{d}$, $\nabla \mb{v}(\mb{x}(t),t)$ is the spatial gradient of velocity at the particle position $\mb{x}$ at time $t$, and $\mb{I}_{d}$ is the identity matrix in the space of the flow \cite{TruesdellRajagopal_IntroToFluids_2000, Haller_DynamicPolarDecompostion_2016}.  Thus, the velocity gradient defines the evolution of the flow map Jacobian.  When deformations are considered over an infinitesimal duration, they are therefore governed by the velocity gradient alone.  

As it will become helpful in subsequent sections, the Euler-Stokes decomposition is introduced, which separates the velocity gradient into a rotational component and a dilatational component  
\begin{equation}
    \nabla\mb{v} = \mb{W} + \mb{D}.  
\end{equation}
The skew-symmetric spin tensor $\mb{W} = \frac{1}{2}(\nabla \mb{v} - \nabla \mb{v}^\top)$ represents the rate of change of material rotation, and the symmetric stretch (or dilatation) tensor $\mb{D} = \frac{1}{2}(\nabla \mb{v} + \nabla \mb{v}^\top)$ represents the rate of stretching in a material element.  One application of these tensors is the computation of vorticity $\mb{\omega}$, which is defined 
\begin{equation}
    \mb{We} = -\frac{1}{2} \bm{\omega}\times \mb{e},\quad \forall \mb{e} \in \R^{d}.
    \label{eq:vorticity}
\end{equation}
Additionally, the principal strain may be computed as the maximum eigenvalue of the strain tensor
\begin{equation}
    \epsilon_1 = \lambda_{max}(\mb{D}),
    \label{eq:principal_stretch}
\end{equation}
and both $\mb{W}$ and $\mb{D}$ define the Q-criterion \cite{Hunt.Moin_QCriteria_1988}
\begin{equation}
    Q = \frac{1}{2} \left(\norm{\mb{W}}_F^2 - \norm{\mb{D}}_F^2\right),
    \label{eq:Q-crit}
\end{equation}
where $\norm{\cdot}_F$ represents the Frobenius norm.

\subsection{The Velocity Gradient in Recent LCS Literature}
Because of this connection, a variety of LCS analyses have been developed involving the velocity gradient.  Serra and Haller use the spin and stretch tensors to construct variational formulae for computing objective Eulerian coherent structures (OECS) \cite{SerraHaller_OECS_2016} which were later implemented as a method for aiding search and rescue efforts \cite{Serra.Haller_SearchAndRescue_2020}.  Along these lines, the instantaneous Lyapunov exponent (iLE) was recently developed using the first three Rivlin-Ericksen tensors to achieve a third-order accurate in time Taylor expansion of the right Cauchy-Green deformation tensor \cite{Nolan.Ross_FTLEInstLimit_2020}.  Additionally, the velocity gradient has been used in the development of the Lagrangian averaged vorticity deviation (LAVD) \cite{Haller.Huhn_LAVD_2016,Haller_DynamicPolarDecompostion_2016} which will be discussed in more depth in section \ref{sec:ellipLCS}. All of these methods tout the advantage of being able to directly use Eulerian velocity fields to construct the metrics.  However, these methods cannot reliably be used when only sparse trajectory information exists, as they require full field information of the flow.  Due to the challenges posed by sparse data (as in the case of oceanic drifters) quasi-objective single-particle metrics were developed that compute measures of stretching and rotation using single-particle velocity and (in the case of rotation) spatially averaged vorticity information \cite{Haller.Encinas-Bartos_Quasi-ObjectiveDiagnostics_2021}.  

Lagrangian tools that use the velocity gradient are not available for sparse data because there are no existing methods that enable the computation of velocity gradients directly from trajectories.  In all cases, velocity field information must be known across the entire domain where gradients are computed.  The particle tracking velocimetry (PTV) community is primarily responsible for developing such tools, since quantities like vorticity and Q are commonly sought from PTV data \cite{Sciacchitano.Schroeder_PTVChallenge_2021, Schanz.Schroder_ShakeTheBox_2016}.  In the PTV framework, velocity gradients are computed by interpolating the scattered velocities of individual particles onto a uniform, Eulerian grid prior applying to a numerical differentiation scheme.  Data-assimilation has been used to enhance the quality of such interpolation schemes \cite{SchroderSchanz_PTVReview_2023}, but its use is associated with large computational expense in terms of storage and computation time.  

This work enables sparse gradient computation through the development of Lagrangian gradient regression (LGR), which identifies velocity gradients by regressing the velocity gradient from the deformation of local trajectories.  This tool was first presented in \cite{Harms.McKeon_LGRISPIV_2023} and is similar to the tool developed independently by Fenelon et al. at the same time \cite{Fenelon.Cattafesta_KinematicDecompositionISPIV_2023}.  The application of LGR to LCS quantities is discussed in sections \ref{sec:hyper_para} and \ref{sec:ellipLCS}.  

\subsection{Estimating the Velocity Gradient using Sparse Trajectories}

Lagrangian gradient regression is founded on the connection between the flow map Jacobian and the velocity gradient as expressed in equation \ref{eq:linearProcessJacobian}.  From it, the velocity gradient can be computed as a function of the flow map Jacobian and its temporal derivative
\begin{equation}
    \nabla \mb{v} = \dfrac{d}{d t}{D\mb{F}}_{t_0}^{t} \left(D\mb{F}_{t_0}^{t}\right)^{-1}.
    \label{eq:velgrad_exact}
\end{equation}
As $t \to t_0$, it is appropriate to approximate $\F \approx \mb{I}_d$.  Therefore, given a time $t$ and a time $\tau = t+\Delta t$ with $\Delta t/ \mathcal{T} << 1$, equation \ref{eq:velgrad_exact} is approximately represented by
\begin{equation}
    \nabla \mb{v} \approx \frac{D\mb{F}_{t}^{\tau} - \mb{I}_{d}}{\Delta t},
    \label{eq:vgradfromF}
\end{equation}
where the result applies at time $t$.  Therefore, an approximation of the velocity gradient is achievable using only trajectory information; it does not require any velocity information or numerical differentiation.  

\subsubsection{Kernel Weighted Least-Squares Regression of $D\mb{F}_{t}^{\tau}$}
For equation \ref{eq:vgradfromF} to be a valid approximation of the velocity gradient, it must first be possible to approximate $D\mb{F}_{t}^{\tau}$ from trajectory information.  To accomplish this, the framework of kernel-weighted least squares regression is employed.  While least-squares regression on particle trajectories has been used in the LCS community (see, for example, Lekien and Ross (2010) \cite{LekienRoss_FTLEUnstructuredMesh_2010}, Raben et al. (2014) \cite{Raben.Vlachos_FTLEonTRPIV_2014}, and Mowlavi et al. (2022) \cite{Mowlavi.Mahadevan_SparseNoisyLCS_2022}), it has not yet been applied for the purpose of velocity gradient estimation, nor has it employed kernel weighting functions.  

The process of kernel weighted least-squares regression of $D\mb{F}_{t}^{\tau}$ begins by identifying a particle of interest at time $\mb{x}(t)$ and selecting all of the surrounding particles $\mb{x}_i(t)$ within an $\epsilon$-neighborhood.  These particles may be native to the flow or artificially seeded, depending on the application.  The distance (typically Euclidean) from all neighboring particles to the central particle at the initial time $\Delta \mb{x}_i(t) = \mb{x}_i(t) - \mb{x}(t)$ is recorded in the matrix of differences at the initial time $\mb{X}_t\in \R^{d\times n}$. These particles are deformed by the flow to their positions at time $\tau$ and the new distances are recorded in a second matrix $\mb{X}_\tau\in \R^{d\times n}$.  The matrices $\mb{X}_t$ and $\mb{X}_\tau$ are constructed for $n$ particles as
\begin{align}
    \mb{X}_t = \begin{bmatrix} \vert & \vert &  & \vert \\ \Delta \mb{x}_1(t) & \Delta \mb{x}_2(t) & \cdots & \Delta \mb{x}_n(t) \\ \vert & \vert &  & \vert \end{bmatrix}, & & \mb{X}_\tau = \begin{bmatrix} \vert & \vert &  & \vert \\ \Delta \mb{x}_1(\tau) & \Delta \mb{x}_2(\tau) & \cdots & \Delta \mb{x}_n(\tau) \\ \vert & \vert &  & \vert \end{bmatrix},
\end{align}
such that $\Delta \mb{x}_i(\tau) = \mb{F}_t^\tau(\mb{x}_i(t)) - \mb{F}_t^\tau(\mb{x}(t))$ represents the distance from the $i$\textsuperscript{th} neighbor particle to the center particle at time $\tau$.  

The deformed positions $\mb{X}_\tau$ can be viewed as a linear mapping from the initial positions $\mb{X}_t$ such that 
\begin{equation}
    \mb{X}_\tau = \mb{A}\mb{X}_t.
\end{equation}
This representation resembles the formulation of dynamic mode decomposition (DMD) \cite{Schmid_DMD_2010, Kutz.Proctor_DMDbook_2016}, as it fits a linear operator to describe the dynamics of material deformation from one instant in time to the next.  The kernel-weighted least-squares regression problem identifies the optimal operator $\mb{A}$ by solving the minimization problem
\begin{equation}
    \mb{A} = \argmin_{\mb{A}} \left(\frac{1}{2}\norm{\mb{K}^\frac{1}{2}\left(\mb{X}_\tau-\mb{AX}_t\right)}_{F}^2 + \frac{\gamma}{2}\norm{\mb{A}}_F^2\right),
    \label{eq:kernel_min}
\end{equation}
where the kernel matrix $\mb{K} \in \R^{n\times n}$ is a design parameter and $\gamma$ is the strength of regularization.  The solution of the optimization is provided by
\begin{equation}
    \mb{A} = \mb{X}_\tau \mb{K} \mb{X}_t^\top \left(\mb{X}_t \mb{K} \mb{X}_t^\top + \gamma n \mb{I}_{d}\right)^{-1},
    \label{eq:regularizedkernelregression}
\end{equation}
which is widely understood in the literature \cite{Bishop_PatternRecognition_2006}.  

Because the operator $\mb{A}$ maps particle positions at time $t$ to their deformed positions at time $\tau$, it represents an approximation of the flow map Jacobian.  Therefore,
\begin{equation}
    D\mb{F}_{t}^{\tau} \approx \mb{A}.
\end{equation}

Selecting a kernel matrix $\mb{K}$ has significant bearing on the quality of the operator identified by equation \ref{eq:regularizedkernelregression}.   Many construction strategies exist.  Valid kernels are known as Mercer kernels \cite{Mercer_Kernel_1909}; as long as $\mb{K}$ is symmetric and positive semi-definite, it is legitimate.  Typically, one defines a kernel matrix according to a kernel function $k(\Delta \mb{x}, \Delta \mb{x}')$, which defines the distance between data in the regression.  

In this paper, $K$ is either set to be the identity matrix, which ascribes equal weight to particles any distance from the center particle, or particles are given weights according to a Gaussian function over the radius.  In the latter case, the kernel function is defined
\begin{equation}
    \mb{k}_{RG}(\Delta \mb{x}_i(t), \Delta \mb{x}_i(t) = \begin{cases}
        \alpha^2 \exp{\left(-\frac{\left(\Delta \mb{x}_i(t)\right)^2}{2 l^2}\right)},& \text{if } i = j\\
        0,              & \text{otherwise}
    \end{cases},
    \label{eq:radialGaussian}
\end{equation}
where the output scaling $\alpha^2$ and the input variance $l^2$ are hyperparameters \cite{Duvenaud_Kernels_2014}.  In practice, allowing $\mb{K} = \mb{I}_n$ is the preferred approach when the particle spacing is already dense relative to the spatial scales of the flow or when the uncertainty of particle trajectories is larger.  By equally weighting over the entire radial domain, the influence of spurious particles is reduced.

\subsubsection{The LGR Algorithm}
When working with sparse, randomly distributed tracer trajectories, LGR is implemented using the procedure outlined in algorithm \ref{alg:lgr}.  Typically, trajectory information is provided as a list of indexed tracers with position histories recorded.  The algorithm first re-orients the data to be indexed in time rather than by particle.  Then, at each time step and for every particle, the nearest neighbors are determined and their relative positions at time $t_i$ and $t_{i+1}$ are recorded.  Equation \ref{eq:regularizedkernelregression} is then used to compute the flow map Jacobian and equation \ref{eq:vgradfromF} is used to compute the velocity gradient.  The result is stored with the tracer at time $t_i$.  
\begin{algorithm}[t]
  \caption{Lagrangian Gradient Regression (LGR) on Sparse Trajectory Data}
  \label{alg:lgr}
  
  \textbf{Input:} Indexed particle trajectories; number of neighbors $n$; kernel function $k(\Delta\mb{x}, \Delta\mb{x})$; regularization parameter $\gamma$. \\
  \textbf{Output:} Velocity gradients recorded along particle trajectories.
  
  \begin{algorithmic}[1]
  \State $N\leftarrow$ number of snapshots recorded.
    \For{each $t_i,\; \forall i \in \{0,\,1,\,\dots,\,N-1\}$ }
      \For{each tracer $\mb{x}(t_i) \in \mathcal{P}$ at $t_i$}
        \State Find nearest neighbors $\mb{x}_j(t_i)$ of $\mb{x}(t_i),\; \forall j\in \{0,\,1,\,\dots,\,n\}$.
        \State Compute kernel matrix $\mb{K}$ using $k(\Delta\mb{x}, \Delta\mb{x})$ with $\Delta\mb{x}_j = \mb{x}_j(t_i)-\mb{x}(t_i)$.
        \State Compute $\Fi$ using equation \ref{eq:regularizedkernelregression}.
        \State Compute $\nabla\mb{v}$ using equation \ref{eq:vgradfromF}.
      \EndFor
    \EndFor
  \end{algorithmic}
\end{algorithm}

\subsection{Results from the Double Gyre}

\begin{figure}[t!]
    \centering
    \includegraphics[trim={2.5cm 1.0cm 2cm 1.0cm},clip, width=1\textwidth]{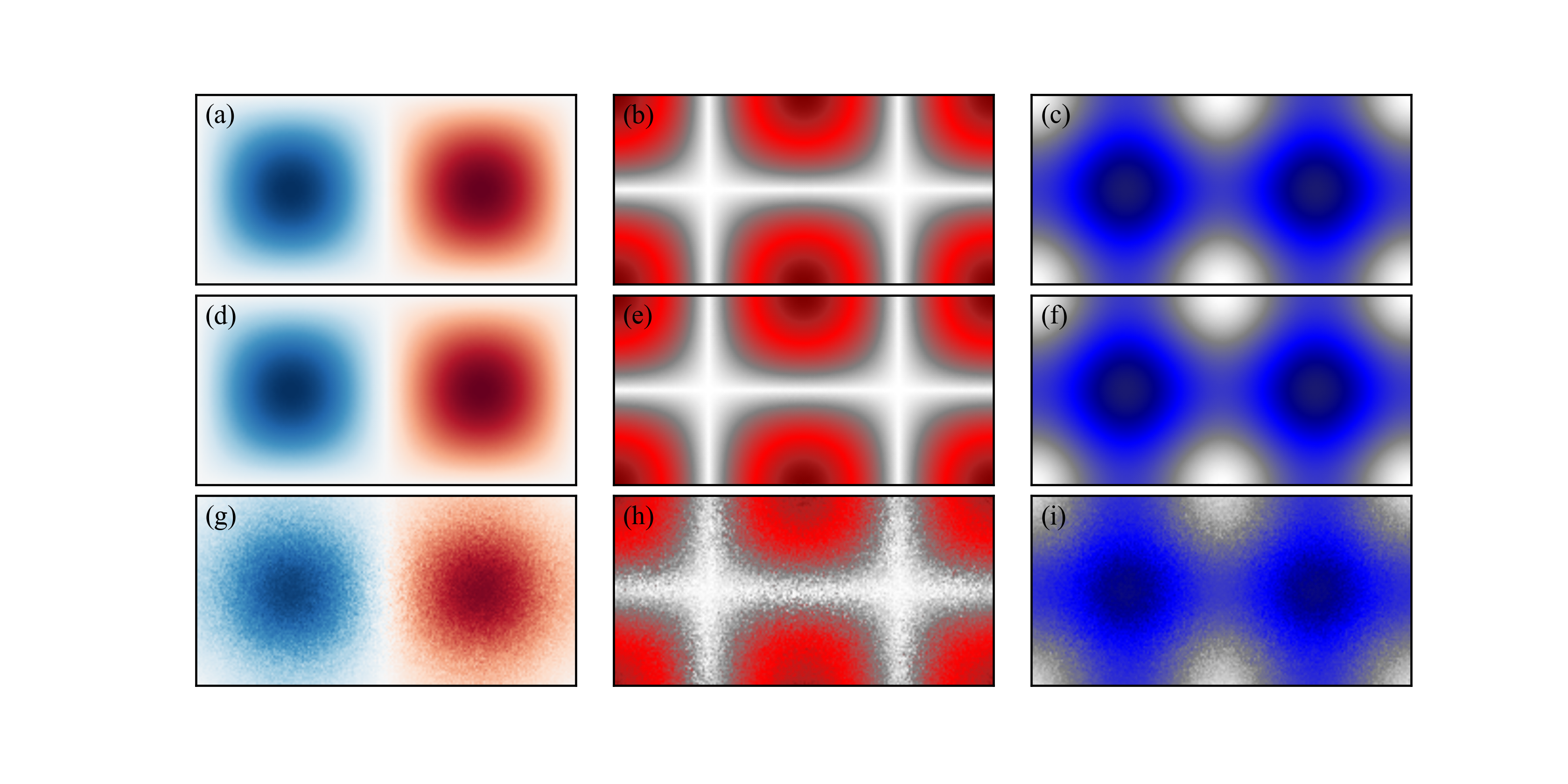}
    \caption{Comparison of quantities derived from the velocity gradient on the double gyre flow at $t=0$ using Lagrangian trajectories computed over $\Delta t = 0.01$.  Vorticity is displayed in the first column, principal strain in the second, and Q-criterion in the third.  (a, b, c) Quantities computed analytically.  (d, e, f) Quantities computed using equation \ref{eq:vgradfromF} on a $200\times 100$ uniform grid using Gaussian weighted regression on 30 random particles within radius $r=0.01$ of the evaluation point. (g, h, i) Same as (d, e, f) with $r=0.5$. }
    \label{fig:FtoL}
\end{figure}

To demonstrate the effectiveness of LGR for computing velocity gradients, velocity, principal strain, and Q are computed both analytically and by LGR on the double gyre flow.  The results are displayed in figure \ref{fig:FtoL}.

To ensure that comparisons are accurate, both the analytical and LGR computations are performed on the same $200\times 100$ uniform grid over the flow domain.  Analytical values are computed directly from equation \ref{eq:doublegyre}, and are displayed in the top row of figure \ref{fig:FtoL} where  figures \ref{fig:FtoL}(a), \ref{fig:FtoL}(b), and \ref{fig:FtoL}(c) represent the vorticity, principal strain, and Q respectively.  LGR computations were performed using the planet-satellite approach where the center particles were assigned to the grid points where analytical measurements were made.  In the second row (figures \ref{fig:FtoL}(d), \ref{fig:FtoL}(e), and \ref{fig:FtoL}(f)), results were computed using regression over 30 particles randomly placed with a uniform distribution within a radius of $r=0.01$ from the test particle.  The kernel matrix $\mb{K}$ was constructed using Gaussian weighting by radius (equation \ref{eq:radialGaussian}) with no regularizer.  In the final row (figures \ref{fig:FtoL}(g), \ref{fig:FtoL}(h), and \ref{fig:FtoL}(i)), LGR was used with the same parameters as in the second row but with sampling radius of $r=0.5$ for.  Thus, the bottom row represents a distribution of particles that is 50 times sparser than in the second row.  

The results displayed in figure \ref{fig:FtoL} demonstrate that LGR is a viable tool for computing velocity gradients directly from particle trajectories.  The fields computed using regression over finely spaced particles (\ref{fig:FtoL}(d)-(f)) are indistinguishable from those computed analytically (\ref{fig:FtoL}(a)-(c)).  The LGR results from sparse distributions (\ref{fig:FtoL}(d)-(f)) are noticeably noisier than the dense computations, which is a result of bias in the sampling distribution at the initial time.  However, the magnitude of the values is correct and the shape and location of the features are still evident.  In cases like this, where initial particle placement is large, using kernel weighting may significantly improve the regression results.  Additionally, if more particles are sampled inside the regression neighborhood, the results exhibit less noise.

%% file: 3_LocallyLinearRegressions.tex
\section{Computing Hyperbolic and Parabolic LCS using LGR\label{sec:hyper_para}}

As discussed in section \ref{sec:geomLCS}, hyperbolic and parabolic LCS are generally computed from the right Cauchy-Green strain tensor $\mb{C}_{t_0}^t$, which is the gram matrix of the flow map Jacobian over the observed time domain.  Because this quantity requires the measurement of material deformation over an extended duration, the sampling density of tracers must be large to ensure that the influence of nonlinearities remains small.  On the other hand, in section \ref{sec:timederivs}, it was discussed that material deformation over short periods of time can be accurately approximated from sparse trajectories since the sampling time $\Delta t \to 0$, preventing the nonlinearities from distorting the result.  In order to extend the framework of LGR to finite observation times---and therefore to the identification of hyperbolic and parabolic LCS---the theory of flow map composition must be considered.    

\subsection{Computing Jacobians using Composition}

The use of flow map compositions for computing $\F$ was pioneered by Brunton and Rowley (2010)  \cite{BruntonRowley_FastComputationFTLE_2010}, Luchtenberg et al.\ (2014) \cite{Luchtenbur.Rowley_UncertaintyPropComposition_2014}, and Brunton (2018) \cite{Brunton_FlowMapComposition_2018}, and has been applied in other studies such as that conducted by Raben et al.\ (2014) \cite{Raben.Vlachos_FTLEonTRPIV_2014}.  The theory stems from the uniqueness and existence properties of the flow map (equation \ref{eq:ivpprops})---particularly the process property
\begin{equation}
    \mb{F}_{t_0}^{t_n}(\mb{x}_0)= \mb{F}_{t_{n-1}}^{t_n}\circ \cdots \circ \mb{F}_{t_{1}}^{t_{2}} \circ \mb{F}_{t_{0}}^{t_{1}}(\mb{x_0}),
    \label{eq:processprop}
\end{equation}
which states that any flow map from time $t_0$ to $t_n$ can be defined as the composition of $n$ intermediate flow maps, as long as there are no gaps in the measurement times.  Applying the chain rule to equation \ref{eq:processprop} yields
\begin{align}
\begin{split}
    D\mb{F}_{t_0}^{t_n}(\mb{x}_0) &= D \left(\mb{\mb{F}}_{t_{n-1}}^{t_n}\circ \cdots \circ \mb{\mb{F}}_{t_{1}}^{t_{2}} \circ \mb{\mb{F}}_{t_{0}}^{t_{1}}(\mb{x_0})\right), \\
    &= D\left(\mb{\mb{F}}_{t_{n-1}}^{t_n} \left(\mb{\mb{F}}_{t_{0}}^{t_{n-1}}(\mb{x}_0)\right)\right) D\left(\mb{\mb{F}}_{t_{n-2}}^{t_{n-1}} \left(\mb{\mb{F}}_{t_{0}}^{t_{n-2}}(\mb{x}_0)\right)\right) \dots  D\left(\mb{\mb{F}}_{t_{0}}^{t_{1}}(\mb{x}_0)\right), \\
    &= D\mb{F}_{t_{n-1}}^{t_{n}} \left(\mb{\mb{F}}_{t_{0}}^{t_{n-1}}(\mb{x}_0)\right)  D\mb{F}_{t_{n-2}}^{t_{n-1}} \left(\mb{\mb{F}}_{t_{0}}^{t_{n-2}}(\mb{x}_0)\right)  \dots  D\mb{F}_{t_{0}}^{t_{1}} \left(\mb{x}_0\right).
\end{split}
\end{align}
Then, recalling from equation \ref{eq:flowmap} that $\mb{x}(t_i) = \mb{F}_{t_0}^{t_{i}}(\mb{x}_0)$, the composition operation can be succinctly stated as
\begin{equation}
    D\mb{F}_{t_0}^{t_n}(\mb{x}_0) = \prod_{i=0}^{n-1}
    D\mb{F}_{t_{i}}^{t_{i+1}} \left( \mb{x}(t_i) \right).
    \label{eq:compeq}
\end{equation}  

It is helpful to make some remarks regarding the composition framework.  First, it is important to notice that equation \ref{eq:compeq} applies to every tracer in the flow for all times along its trajectory regardless of the method used to compute $\Fi$.  Additionally, the important temporal constraint is that computational time domains are consecutive.  Any given interval may progress forward in time, backward in time, or not at all, but all intervals must be connected.  

A second remark is that, while previous methods involving flow map composition achieved their respective goals by performing interpolation to a grid at each time step \cite{BruntonRowley_FastComputationFTLE_2010, Luchtenbur.Rowley_UncertaintyPropComposition_2014, Raben.Vlachos_FTLEonTRPIV_2014}, it is not a requirement for using flow map composition.  As long as calculations are performed consecutively along the particle trajectory, no interpolation is necessary to achieve accurate results.  

Often, when using composition to compute flow map Jacobians, the group of particles identified at the initial time $t_0$ are tracked through intervals to their final positions at time $t_n$.  Computing the flow map Jacobians $\Fi$ at each interval and synthesizing through equation \ref{eq:compeq} to achieve the Jacobian over the full domain $D\mb{F}_{t_0}^{t_n}$ provides exactly the same results as if $D\mb{F}_{t_0}^{t_n}$ were computed using the initial and final times alone.  Thus, composition alone does not solve the problem of sparse identification of LCS.  To overcome this barrier, this work proposes the addition of resampling at each time step.  

\subsection{Composition with Resampling}
\begin{figure}[t!]
    \centering
    \includegraphics[width=1\textwidth]{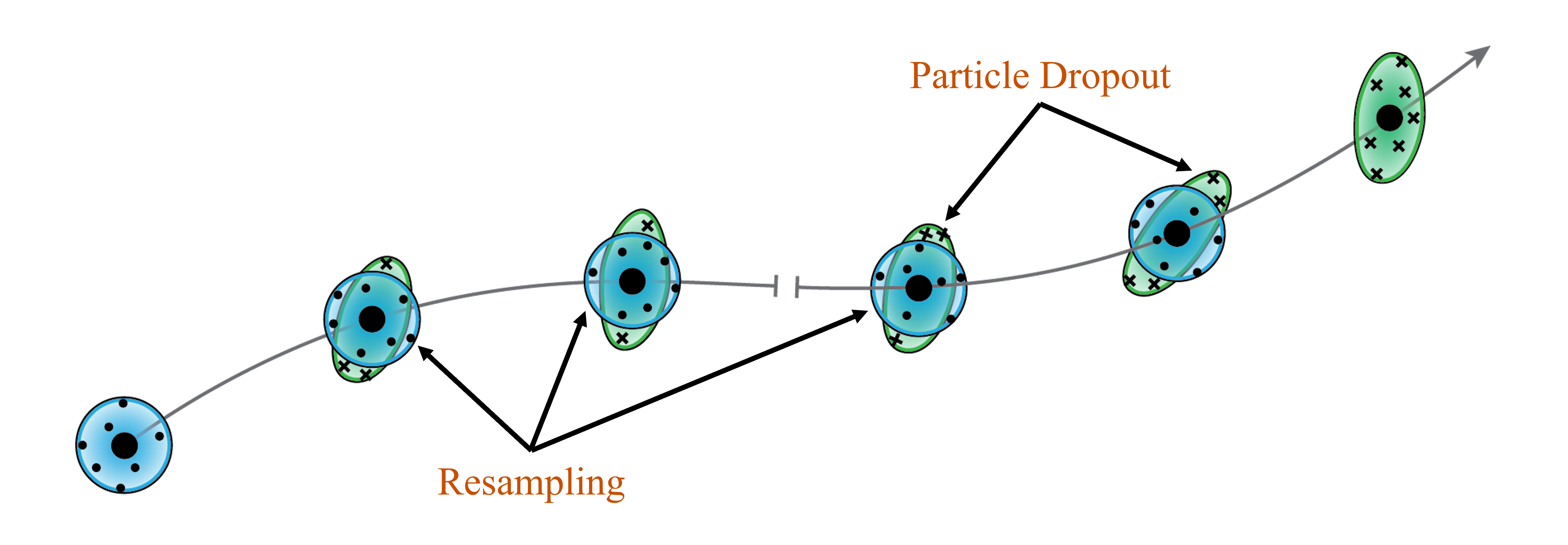}
    \caption{Schematic of the resampling procedure.  At each time step, particles used in regression are resampled to ensure that each regression stays locally linear in time.}
    \label{fig:CWR_schematic}
\end{figure}

The process of composition with resampling is presented as a schematic in figure \ref{fig:CWR_schematic}, which should be considered in comparison with the gradient computation process without resampling from figure \ref{fig:LIT_schematic}.  In the resampling paradigm, tracers which have exceeded a threshold radius from the analyzed trajectory are discarded at each time step and replaced by others that are within closer proximity.  As discussed in the previous section, if the time increment is small enough ($\Delta t/ \mathcal{T} \ll 1$), the material deformation is approximately linear and LGR can be used to fit the short-time Jacobian.  Then, by applying equation \ref{eq:compeq}, the complete Jacobian over time $[t_0,\, t_n]$ can be accurately constructed.  

Depending on the context, resampling may be accomplished using artificial tracers or using those existing in the flow.  If seeding using numerical tracers, one has control over the distribution of tracers in the regression neighborhood.  When computing directly from numerical tracers, some possible approaches include selecting all tracers within a specified radius or selecting the $k$-nearest neighbors to the analyzed trajectory.  In section \ref{sec:randdata}, when the tools developed in this work are demonstrated on sparse, random trajectories, the latter approach is performed.  Regardless of how they are resampled, all tracers used in the regression must persist from one time step to the next to enable regression by equation \ref{eq:regularizedkernelregression}.  However, the neighbor particles may leave the domain outside the step that they are used in regression.  The only particle that must persist the entire duration is the one whose trajectory is analyzed, which is a relaxation from other Jacobian computation methods, since they require all particles to be visible for the entire duration.   

\subsection{Algorithmic Implementation}

Algorithm \ref{alg:ftle} may be used to implement composition for computing flow map Jacobians as described in this section.  It is assumed that the velocity gradients or the intermediate flow map Jacobians are already obtained along each particle trajectory.  These may have been computed by LGR or they might be available through simulated velocity fields or analytical functions.  Given these trajectories with stored intermediate Jacobians, equation \ref{eq:compeq} is applied to the stored Jacobians over some $\Delta t$ at each time step to obtain the long-time flow map Jacobian for each tracer.  The FTLE along the trajectory is then computed using \ref{eq:ftle}.  

\begin{algorithm}[t]
  \caption{FTLE from Sparse Trajectory Data}
  \label{alg:ftle}
  
  \textbf{Input:} Indexed particle trajectories with $\nabla\mb{v}$ available at each time step; observation time $\Delta t$. \\
  \textbf{Output:} FTLE $\sigma_{t_i}^{t_i+\Delta t}$ along each trajectory.
  
  \begin{algorithmic}[1]
  \State $N\leftarrow$ number of snapshots recorded.
    \For{each $t_i,\; \forall i \in \{0,\,1,\,\dots,\,N-1\}$ }  \Comment{At each time step,}
      \For{each tracer $\mb{x}(t_i) \in \mathcal{P}$ at $t_i$}  \Comment{along each trajectory,}
        \State $s \leftarrow i$; $\quad t_n\leftarrow t_i$; $\quad D\mb{F}_{t_i}^{t_s}(\mb{x}(t_i)) \leftarrow \mb{I}_d$.
        \While{$t_n < t_i+\Delta t$} \Comment{perform Jacobian composition.}
          \State $D\mb{F}_{t_i}^{t_s}(\mb{x}(t_i)) \leftarrow D\mb{F}_{t_s}^{t_{s+1}}(\mb{x}(t_s)) D\mb{F}_{t_i}^{t_s}(\mb{x}(t_i))$
          \State $s \leftarrow s+1$; $\quad t_n \leftarrow t_s$
        \EndWhile
        \State $\Delta t_{true} \leftarrow t_n-t_i$; $\quad D\mb{F}_{t_i}^{t_i+\Delta t_{true}}(\mb{x}(t_i)) \leftarrow D\mb{F}_{t_i}^{t_s}(\mb{x}(t_i))$
        \State Compute $\sigma_{t_i}^{t_i+\Delta t_{true}}(\mb{x}(t_i))$ using equation \ref{eq:ftle}. 
      \EndFor
  \EndFor
  \end{algorithmic}
\end{algorithm}

\subsection{FTLE Performance Comparison}
\begin{figure}[t!]
    \centering
    \includegraphics[width=1\textwidth]{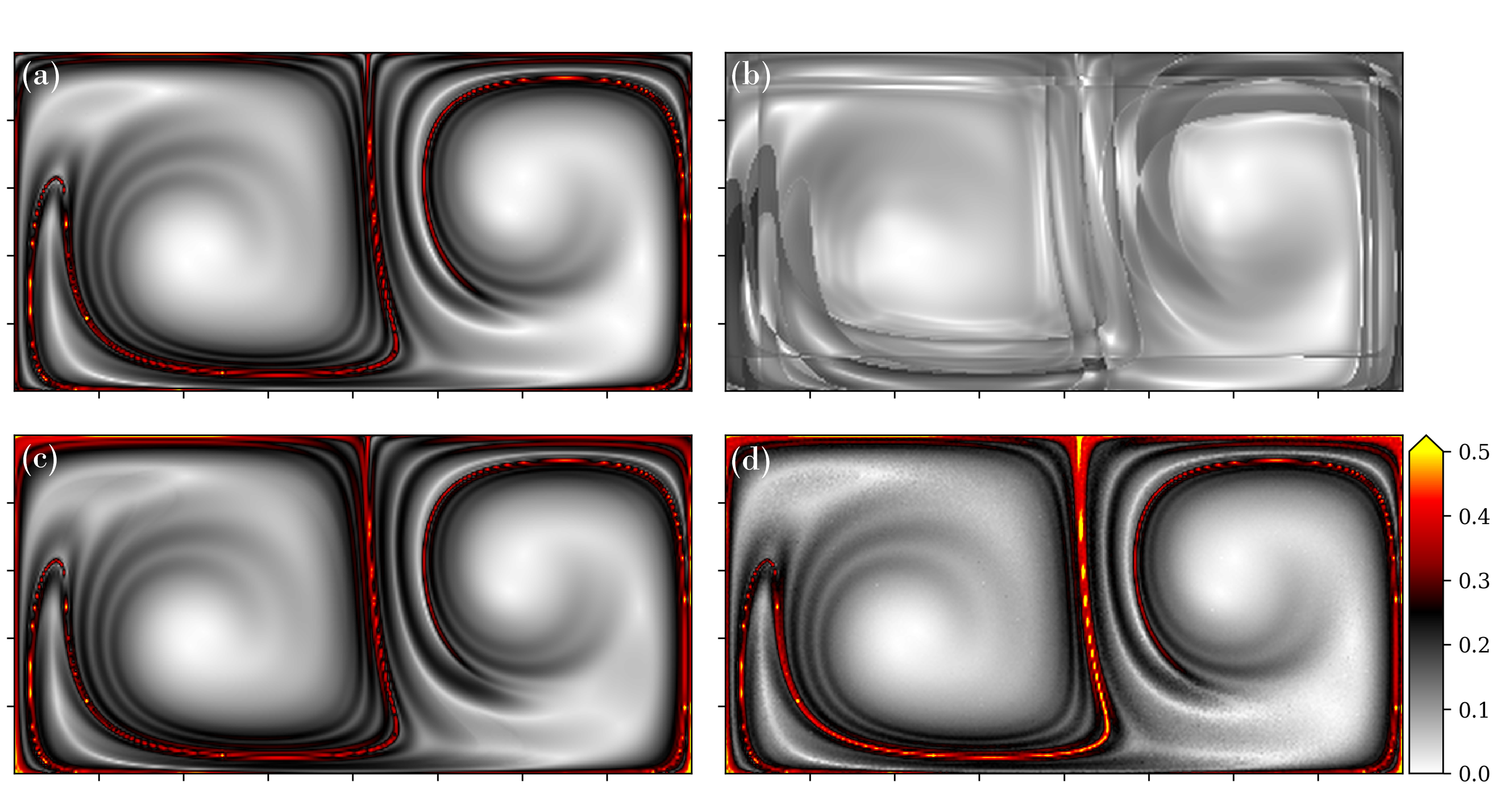}
    \caption{Performance comparison of FTLE computation schemes on the unsteady double gyre flow over 15 time units:  (a) Baseline: FTLE computed using finite-differences with no particle replacement for $\Delta x = 10^{-6}$ by the planet-satellite method (figure \ref{fig:CompSchemes}(c.i)).  (b) Same as (a), but with large $\Delta x = 0.1$.  (c) FTLE computed using finite differences (figure \ref{fig:CompSchemes}(c.i)) with particle replacement (figure \ref{fig:CWR_schematic}) where initial spacing $\Delta x = 0.1$ and replacement time $\Delta t = 0.1$.  (d) Same as (c), but with regression (figure \ref{fig:CompSchemes}(c.ii)) using a Gaussian kernel on the prior distribution.}
    \label{fig:LLR_Perf}
\end{figure}

The effectiveness of LGR with composition for hyperbolic and parabolic LCS detection is demonstrated by example on the double gyre flow.  The parameters outlined with equation \ref{eq:doublegyre} are used and Jacobians are computed over the time domain of $t\in[0,15]$.  The idea of the experiment is to compute a baseline forward FTLE field on the flow using finite differences and small particle spacing and compare it to various results with large initial particle spacing.  The result of the experiment is shown in figure \ref{fig:LLR_Perf} where the color mapping function is kept the same in all frames. 

The baseline FTLE field is displayed in figure \ref{fig:LLR_Perf}(a) and is computed on a $400\times 200$ grid of uniformly spaced interrogation points $\mb{p}$ over the flow domain $[0,2]\times[0,1]$.  To ensure that the same interrogation points can be used between all methods, the planet-satellite computation scheme (figure \ref{fig:CompSchemes}(c)) is used.  Neighboring particles were placed with initial spacing $\Delta \mb{x}=10^{-6}$ according to finite-differences sampling (figure \ref{fig:CompSchemes}(c.i)), which is fine enough to ensure accurate results over the time domain but not small enough as to incur numerical precision error.  For the computation of the Jacobian, central differences were computed using only the initial positions of the particle and their deformed positions at the final time $t=15$.  Note that the ridges are not always smooth and appear aliased in some areas.  This is an artefact of the planet-satellite approach which occurs when $\Delta \mb{p} < 2 \Delta \mb{x}$ and is discussed further in section \ref{sec:randdata}.  

Once again, only the initial and final positions were used to compute the field shown in figure \ref{fig:LLR_Perf}(b), which uses the same computational approach as (a) differing only in the initial particle spacing.  Here $\Delta \mb{x}=0.1$ was used for satellite particle seeding around each interrogation point.  Where initial particles were seeded outside of the flow domain either forward or backward differences were used, neglecting the particle seeded outside of the domain.  The sharp features seen in the field are a numerical artifact that results from the smaller spacing of interrogation points compared to the spacing of satellite particles from which finite-differences are computed $\Delta \mb{p} < \Delta \mb{x}$.  This is why, for example, the relatively fine feature in the lower left of the domain is still faintly visible even though the spacing of regression particles is larger than the feature.  It should be noted that performing FTLE computations in this manner, while technically correct, is unconventional, and has been used here solely for the purpose of direct comparison.

Figures \ref{fig:LLR_Perf}(c) and \ref{fig:LLR_Perf}(d) use LGR with composition to achieve their results.  In both instances, the initial spacing $\Delta \mb{x}=0.1$ is the same as in figure \ref{fig:LLR_Perf}(b).  Satellite particle replacement is employed every $\Delta t = 0.1$ for a total of 150 compositions.  Figure \ref{fig:LLR_Perf}(c) uses structured sampling at each time step and finite differences (figure \ref{fig:CompSchemes}(c.i)) whereas \ref{fig:LLR_Perf}(d) samples $10$ neighboring particles uniformly at random within the radius $r=0.1$ and performs kernel regression (figure \ref{fig:CompSchemes}(c.ii)) with the squared exponential kernel where $\alpha^2 = 1$, $l^2 = \text{var}(\lVert\Delta \mb{x}_i\rVert_2)$ the variance of radial distances of the satellites from the test particle.  A regularizer is set at $\gamma = 10^{-5}$ for numerical stability.  

The results of this experiment demonstrate the effectiveness of the developed methods for accurately reconstructing the flow map Jacobian from sparsely organized particles.  Where the traditional computational strategy of figure \ref{fig:LLR_Perf}(b) does not capture the fine details or the expected ridges seen in the baseline, both approaches using LGR with composition closely match the true values.  The FTLE ridges that are typically used for identifying LCS are clearly present and identifiable, and all of the values in the field are commensurate with the baseline.  In the control case in figure \ref{fig:LLR_Perf}(b) it is not evident that any FTLE ridges exist, and the higher values in the field are significantly diminished.   

The errors in the fields generated using LGR with composition tend to be the highest near the boundaries of the domain, especially near the top boundary, which the particles rotate towards.  This error may be due to issues with numerical sampling when the interrogation region is near the boundary.  Effectively, the distribution of neighboring particles is biased away from the location of the evaluation particle, skewing the results.  In addition, it seems that the principal ridges exhibit slightly larger values than the baseline when using the kernel regression in \ref{fig:LLR_Perf}(d) as opposed to finite-difference resampling in \ref{fig:LLR_Perf}(c).  

\subsection{Single-Trajectory Jacobian Estimation}

Flow map composition also finds utility in circumstances where the velocity gradient is reliably known, as in the case of computing Jacobians on numerically simulated flows.  Rearranging equation \ref{eq:vgradfromF} yields an expression for the flow map Jacobian over sufficiently short durations
\begin{equation}
     D\mb{F}_{t}^{\tau} = \nabla \mb{v} \Delta t + \mb{I}_{d} \text{ as } \Delta t \to 0.
     \label{eq:Ffromvgrad}
\end{equation}
Using this result in tandem with the composition framework (equation \ref{eq:compeq}) allows Jacobians over arbitrary time domains to be computed along a single trajectory.  Because the short-time Jacobians are only a function of the velocity gradient and sampling time, neither finite-differences or regression are used.  

\begin{figure}[t!]
    \centering
    \includegraphics[width=.75\textwidth]{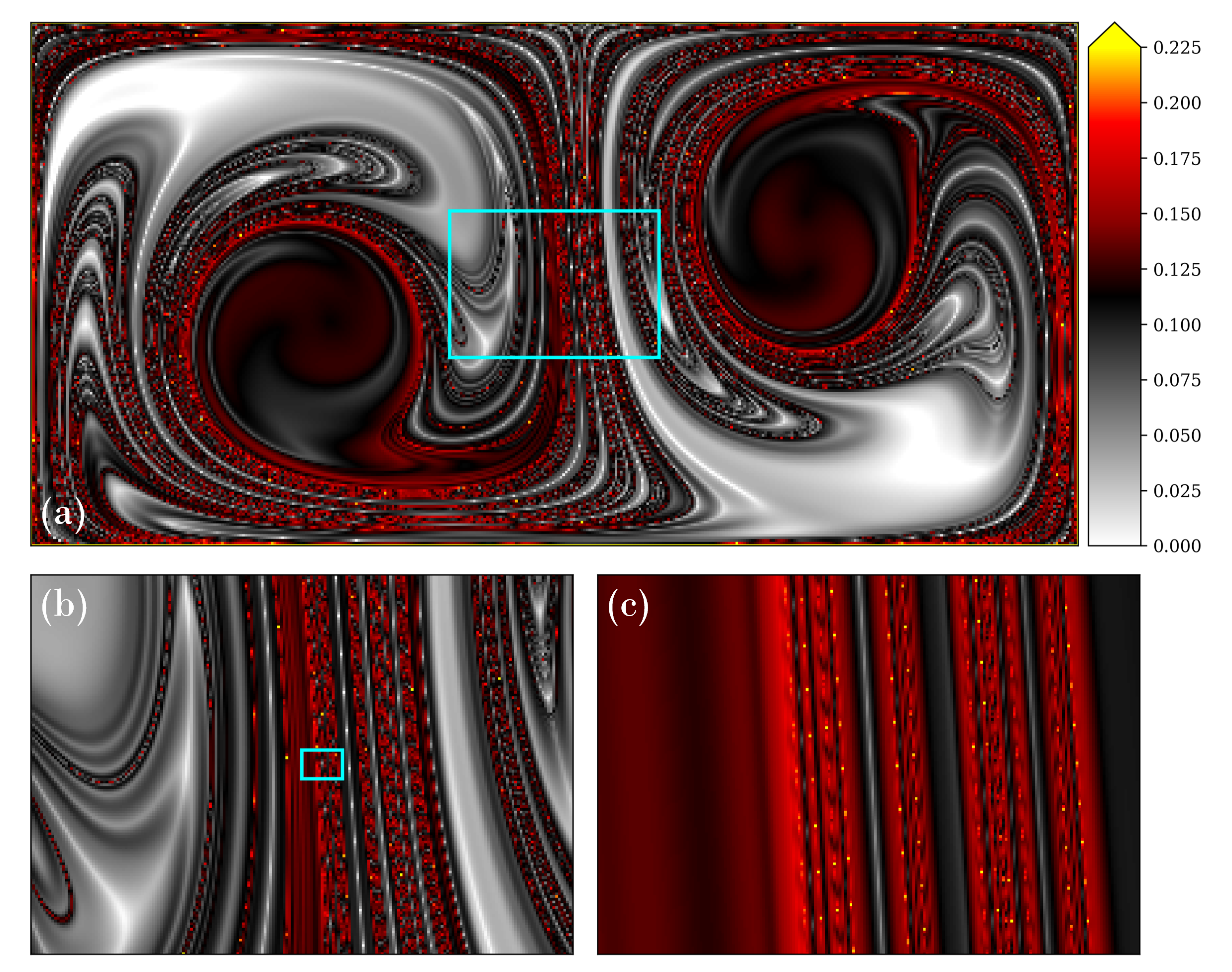}
    \caption{FTLE fields calculated directly from the analytically determined velocity gradient using equations \ref{eq:Ffromvgrad} and \ref{eq:compeq}.  Computations are performed over $t\in[0,\,45]$ with $\Delta t = 0.25$ on three domains of varying density.  (a) The full domain with uniform evaluation particle spacing at $\Delta \mb{p}=0.005$.  (b) A reduced domain with uniform evaluation particle spacing at $\Delta \mb{p}=0.002$. (c) A further reduced domain with uniform evaluation particle spacing at $\Delta \mb{p}=0.00015$}
    \label{fig:LtoF}
\end{figure}

The characteristics of this approach may be seen more clearly by an example, which is presented in figure \ref{fig:LtoF}.  Using the analytically determined $\gradv$ to compute $\Frel$ by equation \ref{eq:Ffromvgrad}, the FTLE field is computed on the double gyre flow over the time domain $[0,\,45]$ with $\Delta t=0.25$ between snapshots.  Computations are performed on three spatial domains that are decreasing in size to illustrate properties of the method.  The domain in figure \ref{fig:LtoF}(a) covers the entire flow and is discretized with particles on a uniform mesh where $\Delta \mb{p}=0.005$.  The domain in figure \ref{fig:LtoF}(b) is the subset of the first indicated by the cyan box in \ref{fig:LtoF}(a), and has uniform particle spacing of $\Delta \mb{p}=0.002$.  The domain in figure \ref{fig:LtoF}(c) is outlined by the cyan box in \ref{fig:LtoF}(b) and has uniform particle spacing of $\Delta \mb{p}=0.00015$.

An advantage of this approach is that particle trajectories do not need to be structured for the computation to be successful.  They can be ordered in any way that is convenient, and they do not need to be integrated simultaneously.  This may allow for optimal sampling strategies similar to those considered by Lekien and Ross (2010) \cite{LekienRoss_FTLEUnstructuredMesh_2010} to reduce computation time.  Another advantage of the method is that the observed features remain sharp when computed over long time domains.  In the standard approach, if the time domain increases, the particle spacing must decrease to avoid the distorting influence of nonlinear effects.  Notably, however, the ridges in figure \ref{fig:LtoF} appear aliased at every spatial resolution.  This is due to the infinitesimal thickness of hyperbolic structures, and it was recently noted by Badza et al. \cite{Badza.Balasuriya_LCSSensitivity_2023} that methods which detect full-dimensional coherent regions are more robust than those, like FTLE, which detect lower-dimensional flow barriers.  Therefore, this approach may find more utility in computing elliptic LCS than hyperbolic LCS.

%% file: 4_EllipticLCS.tex
\section{Computing Elliptic LCS from Trajectories using LGR \label{sec:ellipLCS}}

Identifying elliptic LCS can be thought of as a Lagrangian approach to defining vortex boundaries.  The first of such analyses was accomplished by Haller and Beron-Vera in 2012 \cite{HallerBeron-Vera_GeodesicTheory2D_2012, HallerBeron-Vera_CoherentVortices_2013} using a variational approach that identified closed bands of fluid that demonstrated minimal variability in their average straining.  Later, Farazmand and Haller defined polar LCS as closed and connected material surfaces (loops in 2-D or tubes in 3-D) of the polar rotation angle (PRA) $\theta_{t_0}^t$ \cite{FarazmandHaller_PolarRotationAngle_2016}.  LCS defined from the polar rotation angle, however, are only objective in two dimensions and can be difficult to interpret since  $\theta_{t_0}^t \in [0,\,2\pi)$.  In response to this, Haller derived a generalized polar decomposition \cite{Haller_DynamicPolarDecompostion_2016} which enabled rotationally coherent Lagrangian vortices to be identified by calculating the Lagrangian-averaged vorticity deviation (LAVD) \cite{Haller.Huhn_LAVD_2016}.  In all of these instances, full flow-field information is required and the LAVD in particular necessitates that velocity gradients can be reliably computed over the spatial domain.  Thus, these quantities have not yet been directly computed from tracer trajectories alone.  

In this section, it is shown how LGR enables the computation of elliptic LCS metrics such as the LAVD directly from sparse particle trajectories.  By computing velocity gradients along Lagrangian trajectories, it is possible to compute elliptic LCS without any field information.  

\subsection{Background Theory for Elliptic LCS}
Defining elliptic LCS from sparse trajectory data alone depends on theoretical results from prior work in elliptic LCS theory.  In particular, the polar decomposition and the dynamic polar decomposition must be discussed, as should the definitions of the Lagrangian-averaged vorticity deviation (LAVD) and the dynamic rotation angle (DRA).    

\subsubsection{The Polar Decomposition}
The polar decomposition is essential to understanding the rotational properties Lagrangian tracers and therefore also for developing tools assessing elliptic LCS.  The polar decomposition states that $\F$ has the unique left and right decompositions
\begin{equation}
    \F = \Rpd \Upd = \Vpd \Rpd,
\end{equation}
where the rotation tensor $\Rpd$ is proper orthogonal and the right and left stretch tensors $\Upd$ and $\Vpd$ are symmetric, positive definite \cite{Gurtin.Anand_ContinuumMechanics_2010}.  Typically, $\Rpd$ is interpreted as a solid body rotation of a material element from time $t_0$ to time $t$ and $\Upd$ and $\Vpd$ are associated with the right and left Cauchy-Green strain tensors by $\C = (\F)^\top \F = (\Upd)^2$ and $\B = \F(\F)^\top= (\Vpd)^2$ and therefore represent the stretching behavior over the interval.  It may be noted that $\Upd$ and $\Vpd$ are the right and left singular vectors of the flow map Jacobian $\F$. 

Since the rotation tensor is proper orthogonal, it has an eigenvalue that is equal to unity.  The orientation of the eigenvector associated with the unit-valued eigenvalue is called the polar rotation angle $\theta_{t_0}^t$. 
A variety of strategies have been developed to compute $\theta_{t_0}^t$ for both 2 and 3 dimensions, including work done by Guan-Suo \cite{Guan-Suo_DeterminingRotationTensor_1998} and Farazmand and Haller \cite{FarazmandHaller_PolarRotationAngle_2016}, the latter of which was used for determination of polar LCS.  

Using $\theta_{t_0}^t$ or $\Rpd$ to identify LCS, however, is limited by some of the properties that the quantities possess.  For instance, the polar LCS defined by level-sets of the $\theta_{t_0}^t$ are only frame invariant in planar flows \cite{Haller.Huhn_LAVD_2016}.  In three dimensions, they depend on the reference frame \cite{FarazmandHaller_PolarRotationAngle_2016}.  Additionally, qualitative consideration of the PRA field is difficult since the values of $\theta_{t_0}^t$ are limited between 0 and $2\pi$.    

Another analytical challenge stems from the dynamical inconsistency of $\Rpd$ \cite{Haller_DynamicPolarDecompostion_2016}.  The evolution of the rotation tensor $\Rpd$ is governed by the complex relationship 
\begin{equation}
    \dfrac{d}{d t} \Rpd = \left(\mb{W}(\mb{x}(t),t) - \frac{1}{2}\left(\dfrac{d}{d t} \Upd (\Upd)^{-1} - (\Upd)^{-1}\dfrac{d}{d t} \Upd \right) (\Rpd)^\top \right) \Rpd
    \label{eq:Rpd_evolution}
\end{equation}
(see Truesdell and Rajagopal \cite{TruesdellRajagopal_IntroToFluids_2000}).  Each subsequent state depends not only on the current time $t$ and the previous state $\Rpd$, but also on $t_0$ through $\Upd$.  This memory effect prevents iterative computation of $\Rpd$ as a non-autonomous differential equation.  Because of this, the rotation tensor is dynamically inconsistent, which implies that, in general,
\begin{equation}
    \Rpd \neq \mb{R}_s^t \mb{R}_{t_0}^s
\end{equation}
    for two connected intervals $[t_0,\,s]$ and $[s,\,t]$ \cite{Haller_DynamicPolarDecompostion_2016}.  Therefore, the polar rotation over the full time domain cannot be produced through temporal superposition of intermediate rotation angles, 
\begin{equation}
    \theta_{t_0}^t \neq \theta_{s}^t + \theta_{t_0}^s.
\end{equation}
Since the PRA is dynamically inconsistent, there is no intuitive link between the polar rotation angle and the evolving vorticity of the flow field.  These shortcomings are addressed in detail in the appendices of Haller et al. 2016 \cite{Haller.Huhn_LAVD_2016}.

\subsubsection{The Dynamic Polar Decomposition}
The dynamic polar decomposition (DPD) addresses the difficulties posed by the classic polar decomposition.  Developed by Haller in 2016 \cite{Haller_DynamicPolarDecompostion_2016}, the DPD splits any tensor defined by a linear process into a rotational process with zero rate of strain and an irrotational process with no vorticity.  Specifically, the decomposition is defined 
\begin{equation}
    \F = \DG{O} \DG{M},
\end{equation}
where $\DG{O}$ is the proper orthogonal dynamic rotation tensor and $\DG{M}$ is the right dynamic stretch tensor.  Both $\DG{O}$ and $\DG{M}$ can be expressed as linear initial value problems on purely rotational and purely straining flows respectively
\begin{subequations}
\begin{align}
    \dfrac{d}{d t}\DG{O} &= \mb{W} \DG{O} \label{eq:O_evolution}\\
    \dfrac{d}{d t}\DG{M} &= \left((\DG{O})^{-1}\mb{D} \DG{O}\right)\DG{M}, 
\end{align}
\end{subequations}
where $\mb{O}_{t_0}^{t_0} = \mb{M}_{t_0}^{t_0} = \mb{I}_{d}$ Unlike the classic rotation tensor $\Rpd$, the dynamic rotation tensor $\DG{O}$ is dynamically consistent and therefore satisfies
\begin{equation}
    \DG{O} = \mb{O}_s^t \mb{O}_{t_0}^s
\end{equation}
for any connected intervals $[t_0,\,s]$ and $[s,\,t]$.  However, because it is dependent on the spin tensor $\mb{W}$, it is not objective.

Haller further showed \cite{Haller_DynamicPolarDecompostion_2016} that the dynamic rotation tensor $\DG{O}$ can itself be decomposed into a relative rotation tensor $\DG{\Phi}$ and a mean rotation tensor $\DG{\Theta}$ by 
\begin{equation}
    \DG{O}(\mb{x}_0) = \DG{\Phi} \DG{\Theta}.
\end{equation}
Both the relative and mean rotation tensors are solutions of the initial value problems
\begin{subequations}
\begin{align}
    \dfrac{d}{d t}\DG{\Phi} &= \left(\mb{W}(\mb{x}, t) - \xoverline{\mb{W}}(t)\right) \DG{\Phi}\\
    \dfrac{d}{d t}\DG{\Theta} &= \left((\DG{\Phi})^{-1}\xoverline{\mb{W}}(t) \DG{\Phi}\right)\DG{\Theta},
\end{align}
\end{subequations}
where $\mb{\Phi}_{t_0}^{t_0} = \mb{\Theta}_{t_0}^{t_0} = \mb{I}_{d}$ and $\xoverline{\mb{W}}(t)$ represents the spatial average of $\mb{W}$ over all observed particles.  The relative deformation tensor $\DG{\Phi}$ is dynamically consistent and objective in two dimensions, while the mean deformation tensor $\DG{\Theta}$ is not dynamically consistent because it exhibits the memory effects discussed above \cite{Haller_DynamicPolarDecompostion_2016}. All together, the dynamic polar decomposition has the form
\begin{equation}
    \F = \DG{\Phi} \DG{\Theta} \DG{M}.
\end{equation}

\subsubsection{Metrics for Elliptic LCS Identification}

Using the analytical framework of the dynamic polar rotation, Haller \cite{Haller_DynamicPolarDecompostion_2016} presented the LAVD and the DRA as Lagrangian metrics for elliptic LCS identification.  

Because the relative rotation tensor $\DG{\Phi}$ of the dynamic polar decomposition is dynamically consistent, the angle that it sweeps around its own axis of rotation---the so-called intrinsic rotation angle $\psi_{t_0}^t$---is also dynamically consistent \cite{Haller_DynamicPolarDecompostion_2016, Haller.Huhn_LAVD_2016}.  Therefore,
\begin{equation}
    \psi_{t_0}^t = \psi_{s}^t + \psi_{t_0}^s
\end{equation}
on connected intervals.  

The intrinsic rotation angle (IRA) is computed using
\begin{equation}
    \psi_{t_0}^t(\mb{x}_0) = \frac{1}{2} \int_{t_0}^t |\mb{\omega}(\mb{x}(\tau; t_0, \mb{x}_0), \tau) - \xoverline{\mb{\omega}}(\tau)| d\tau.
    \label{eq:ira}
\end{equation}
From this, the Lagrangian-averaged vorticity deviation (LAVD) is defined as 
\begin{equation}
    \text{LAVD}_{t_0}^t = 2 \psi_{t_0}^t.
    \label{eq:LAVD}
\end{equation}
Therefore, the LAVD can be understood conceptually as twice the amount of rotation that a material element experiences in the analyzed duration about its own intrinsic axis of rotation.  

The dynamic rotation angle (DRA) $\varphi_{t_0}^t$, on the other hand, represents the amount of rotation that a fluid element experiences relative to the observer.  It is defined as 
\begin{equation}
    \varphi_{t_0}^t(\mb{x}_0; \mb{g}) = -\frac{1}{2} \int_{t_0}^t \mb{\omega}(\mb{x}(\tau),\tau) \cdot \mb{g}(\mb{x}(\tau),\tau) d\tau,
    \label{eq:DRA}
\end{equation}
where $\mb{g}$ is an axis family related to the observer around which rotations are measured, the DRA is the angle generated by the dynamic rotation tensor $\DG{O}$ \cite{Haller_DynamicPolarDecompostion_2016}.  Because of the dependence on the observer, the dynamic rotation angle is not objective.  It is, however, dynamically consistent, and provides information about the direction of rotation where the LAVD does not.  

\subsection{Computing LAVD and DRA from Trajectories}
While LAVD and DRA are useful for analyzing the rotational properties of flows over finite times, the current methods for computing them are not applicable to sparse data because they depend on velocity gradients, which are often unreliable and for which no Lagrangian algorithms yet exist.  However, in section \ref{sec:timederivs} Lagrangian gradient regression (LGR) was introduced as a tool for calculating velocity gradients using discrete trajectories defined over short time intervals.  Therefore, using LGR, it is now possible to compute the LAVD and DRA directly from sets of sparse trajectories.  

The process for computing LAVD and DRA using LGR involves computing the short-time flow map Jacobian $\Fi$ for every interval $[t_i,\, t_{i+1}]$ in $[t_0,\, t_n]$ by regression as in equation \ref{eq:regularizedkernelregression}.  From this, the velocity gradient $\nabla \mb{v}$ is computed at every $t_i$ by equation \ref{eq:vgradfromF}.  The vorticity $\mb{\omega}(\mb{x}_i, t_i)$ can then be defined using equation \ref{eq:vorticity}.  For LAVD it is necessary to compute the vorticity deviation $\mb{\omega}'(\mb{x}_i, t_i) = \left \lvert\mb{\omega}(\mb{x}_i, t_i) - \overline{\mb{\omega}}(t_i)\right\rvert$, where the spatial average of vorticity is computed by averaging over all of the particles observed in the domain 
\begin{equation}
    \overline{\mb{\omega}}(t_i) \approx \frac{1}{N} \sum_{\mb{x}_i \in \mathcal{P}} \omega(\mb{x}_i, t_i), 
    \label{eq:spatavg_vort}
\end{equation}
where $\mathcal{P}$ is the set of all $N$ particles observed at time $t_i$.  Then, the IRA and the DRA can be approximated using the equations
\begin{subequations}
\begin{align}
    \psi_{t_0}^{t_n} &= \frac{1}{2}\sum_{i=1}^{n} \left \lvert\mb{\omega}(\mb{x}_i, t_i) - \overline{\mb{\omega}}(t_i)\right\rvert \Delta t_i \label{eq:approxLAVD} \\
    \varphi_{t_0}^{t_n} &= -\frac{1}{2}\sum_{i=1}^{n} \mb{\omega}(\mb{x}_i, t_i) \cdot \mb{g}(\mb{x}(t_i), t_i) \Delta t_i \label{eq:approxDRA}
\end{align}
\end{subequations}
as $n\to \infty$ over a finite time domain.  LAVD is then computed by equation \ref{eq:LAVD}.  Practically speaking, the dependence of the DRA on the observer through the sequence of vectors $\mb{g}(\mb{x}(t_i), t_i)$ renders its computation more challenging in three dimensions.  However, since the vorticity is always normal to the flow plane in 2-D flows, it is a useful metric in that context.  

\subsubsection{Algorithmic Implementation}

As in algorithm \ref{alg:ftle}, computing LAVD and DRA from sparse trajectories requires that velocity gradient information be known at each intermediate position along every particle trajectory.  In the context of sparse trajectory data, this is accomplished using LGR as a prior step to implementation.  Assuming that this requirement is met, LAVD and DRA are computed using algorithm \ref{alg:lavd_dra}, where at each time step and for each tracer, vorticity and vorticity deviation are computed at each interval in the observed time domain and summed according to equations \ref{eq:approxLAVD} and \ref{eq:approxDRA} to achieve the LAVD and DRA respectively.  
\begin{algorithm}[t]
  \caption{LAVD and DRA from Sparse Trajectory Data}
  \label{alg:lavd_dra}
  
  \textbf{Input:} Indexed particle trajectories with $\nabla\mb{v}$ available at each time step; observation time $\Delta t$. \\
  \textbf{Output:} LAVD $\text{LAVD}_{t_i}^{t_i+\Delta t}$ and DRA $\varphi_{t_i}^{t_i+\Delta t}$ along each trajectory.
  
  \begin{algorithmic}[1]
  \State $N\leftarrow$ number of snapshots recorded.
    \For{each $t_i,\; \forall i \in \{0,\,1,\,\dots,\,N-1\}$ }  \Comment{At each time step,}
      \For{each tracer $\mb{x}(t_i) \in \mathcal{P}$ at $t_i$}  \Comment{along each trajectory,}
        \State $s \leftarrow i$; $\quad t_n\leftarrow t_i$; $\quad \psi_{t_i}^{t_s}(\mb{x}(t_i)) \leftarrow 0$; $\quad \varphi_{t_i}^{t_s}(\mb{x}(t_i)) \leftarrow 0$.
        \While{$t_n < t_i+\Delta t$} \Comment{accumulate rotation angles.}
          \State Compute vorticity $\mb{\omega}(\mb{x}_s, t_s)$ by equation \ref{eq:vorticity}.
          \State Compute spatial averaged vorticity $\overline{\mb{\omega}}(t_s)$ by equation \ref{eq:spatavg_vort}.
          \State $\psi_{t_i}^{t_s}(\mb{x}(t_i)) \leftarrow \psi_{t_i}^{t_s}(\mb{x}(t_i)) +  \frac{1}{2}\left \lvert\mb{\omega}(\mb{x}_s, t_s) - \overline{\mb{\omega}}(t_s)\right\rvert (t_s - t_i)$
          \State $\varphi_{t_i}^{t_s}(\mb{x}(t_i)) \leftarrow \varphi_{t_i}^{t_s}(\mb{x}(t_i)) + \frac{1}{2}\mb{\omega}(\mb{x}_s, t_s)\cdot \mb{g}(\mb{x}(t_s), t_s) (t_s - t_i)$
          \State $s \leftarrow s+1$; $\quad t_n \leftarrow t_s$
        \EndWhile
        \State $\Delta t_{true} \leftarrow t_n-t_i$; $\quad \text{LAVD}_{t_i}^{t_i+\Delta t_{true}}(\mb{x}(t_i)) \leftarrow 2 \psi_{t_i}^{t_s}(\mb{x}(t_i))$; $\quad \varphi_{t_i}^{t_i+\Delta t_{true}}(\mb{x}(t_i)) \leftarrow \varphi_{t_i}^{t_s}(\mb{x}(t_i))$
      \EndFor
  \EndFor
  \end{algorithmic}
\end{algorithm}

\subsubsection{Connection to the Polar Rotation Angle}
The DRA is connected to the polar rotation angle when computed over small time domains, and can therefore be computed by an alternate method.  Considering equation \ref{eq:Rpd_evolution} in the limit as $t\to t_0$ and comparing with equation \ref{eq:O_evolution}, it is seen that the rotation tensor and the dynamic rotation tensor are identical \cite{Haller_DynamicPolarDecompostion_2016}
\begin{equation}
    \dfrac{d}{dt}\left.\Rpd \right \rvert_{t_0 = t} = \dfrac{d}{dt}\left.\DG{O} \right \rvert_{t_0 = t} = \mb{W}(\mb{x}_0, t_0).
\end{equation}
Therefore, the polar rotation angle $\theta_{t_0}^t$ is identical to the dynamic rotation angle $\varphi_{t_0}^t$ for infinitesimally small deformations
\begin{equation}
    \left.\theta_{t_0}^t \right \rvert_{t_0 = t} = \left.\varphi_{t_0}^t \right \rvert_{t_0 = t} = -\frac{1}{2} \mb{\omega}(\mb{x}_0,t_0) \cdot \mb{g}(\mb{x}_0, t_0),
\end{equation}
and is proportional to the vorticity at $t_0$ as seen by the observer.  Thus, in two dimensions, the DRA can also be represented as the sum of intermediate polar rotation angles computed between each snapshot (assuming that rotations between snapshots remain smaller than $2\pi$).  

\subsection{Demonstration on the Double Gyre}
\begin{figure}[t!]
    \centering
    \includegraphics[width=1\textwidth]{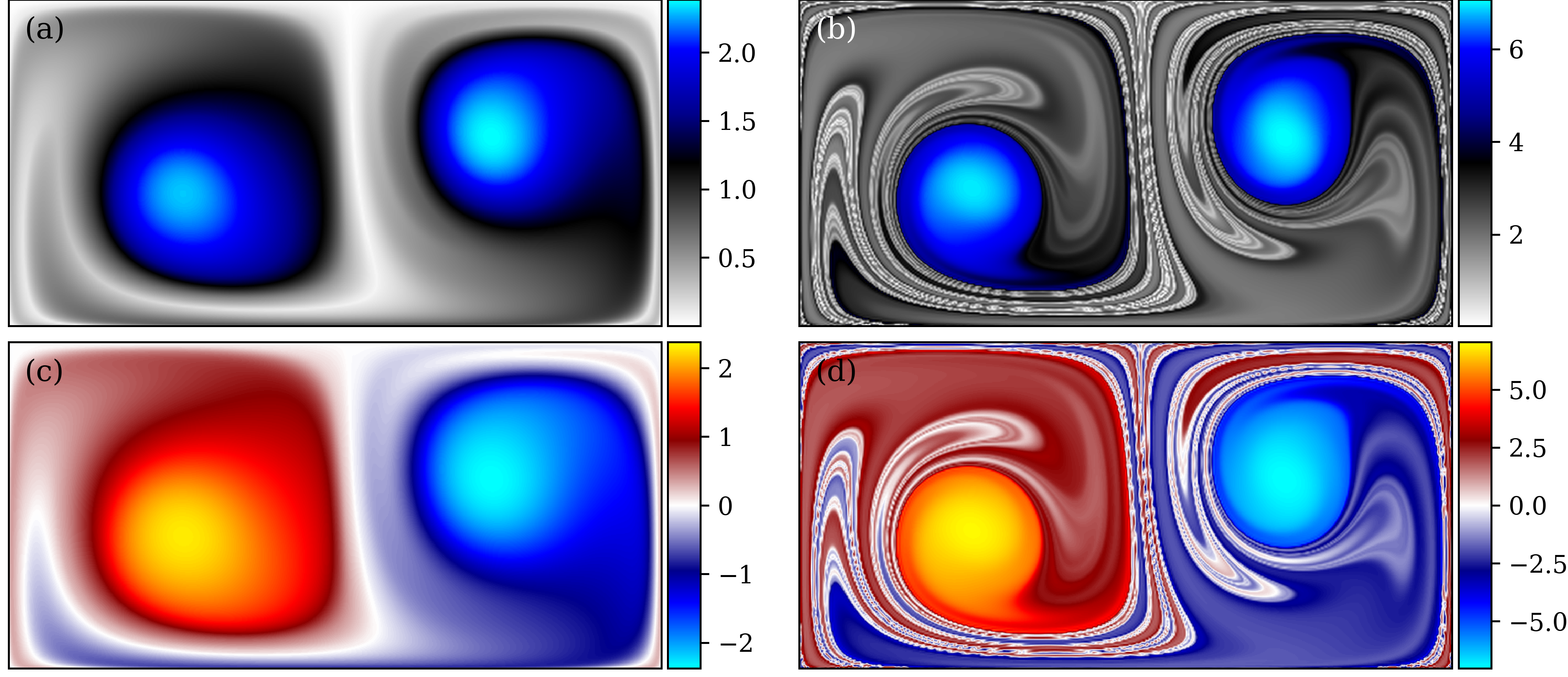}
    \caption{Examples of the Lagrangian metrics for elliptic LCS detection.  Fields in the left column are computed on $t\in [0,15]$ with $\Delta t = 0.1$ on a uniform grid with $\Delta \mb{p}_0=0.005$ using equation \ref{eq:Ffromvgrad}.  Fields in the right column are the same except for $t\in [0,45]$.  (a,b) Intrinsic rotation angle $\psi_{t_0}^{t}$ measured in complete rotations.  (c,d) Dynamic rotation angle $\varphi_{t_0}^t$ measured in complete rotations.}
    \label{fig:EllipMetrics}
\end{figure}

A brief demonstration of the elliptic measures developed in this section is presented on the double gyre flow.  The intrinsic rotation angle (half the LAVD) and the dynamic rotation angle are computed over $t\in[0,\,15]$ in figure \ref{fig:EllipMetrics}(a) and \ref{fig:EllipMetrics}(c) respectively, and again over $t\in[0,\,45]$ in figure \ref{fig:EllipMetrics}(b) and \ref{fig:EllipMetrics}(d).  The relative Jacobians $\Frel$ are computed using equation \ref{eq:Ffromvgrad} with $\Delta t = 0.1$.  Computations are performed along trajectories with initial positions uniformly spaced over the entire domain with $\Delta \mb{p}_0 = 0.005$.  The color mapping is scaled to represent complete rotations of a fluid element, where a positive value represents counterclockwise rotation in figure \ref{fig:EllipMetrics}(c) and (d).    

A few observations should be made regarding figure \ref{fig:EllipMetrics}.  First, considering the intrinsic rotation angle in (a) and (b), the vortex cores (peaks in the $\psi_{t_0}^t$ field) are clearly visible as bright blue patches on either side of the domain.  It is notable that the peaks are not centered with the peak in $|\omega_z|$ at any time in the computation, which is consistent with LCS and LAVD literature \cite{Haller.Huhn_LAVD_2016}.  Additionally, the valleys of the $\psi_{t_0}^t$ field correspond to regions of maximal stretching.

Next the DRA in \ref{fig:EllipMetrics}(c) and (d) is considered.  It should first be noted that the magnitude of rotations expressed in the color bar is consistent with the magnitude of rotation seen in the intrinsic rotation fields. In both cases, the flow experiences just over 2.25 rotations in $t\in[0,\,15]$ and close to 7 in $t\in[0,\,45]$.  Here it is also worthwhile to consider where the field is equal to zero.  It is seen that these contours align with the ridges of the FTLE field in most (but not all) areas.  Thus, the regions where average rotation is close to zero is highly correlated with the regions where there is the most stretching in a flow as indicated by the FTLE.  However, since the DRA is not objective, the similarities between $\varphi_{t_0}^t$ and $\psi_{t_0}^t$ may be fewer in other, more complex flows.

%% file: 5_RandomParticleFields.tex
\section{Application of LGR to Sparse Data \label{sec:randdata}}

Lagrangian gradient regression was introduced as a tool for identifying short-time flow-map Jacobians and velocity gradients in section \ref{sec:timederivs}.  It was then used to compute hyperbolic and parabolic LCS metrics in section \ref{sec:hyper_para} and elliptic LCS metrics in section \ref{sec:ellipLCS}.  The methods were designed with intent to provide accurate results on sparse, randomly distributed trajectory data.  In this section, LGR is applied to fields of random particles with varied density and the quality of the results is evaluated.  It will be shown that it effectively computes velocity gradients with no field information or differentiation, outperforms the standard method when computing FTLE on sparse trajectories, and extracts elliptic fields directly from trajectory data.  

The numerical experiments involve tracers of varying density propagated through the double gyre flow (equation \ref{eq:doublegyre} with the associated parameters) on the spatial domain $[0,\,2] \times [0,\,1]$ from $t_0=0$ to $t=15$.  This allows for direct comparison to results presented earlier in the paper.  Neighboring tracers are resampled every $\Delta t = 0.1$.  Regression between snapshots is performed using $k=15$ nearest neighbors with the radial-Gaussian weighting function (equation \ref{eq:radialGaussian}).  When there are fewer than 1000 particles in the frame, radial basis function (RBF) interpolation with a multiquadric kernel function is used to compute particle values on a $200 \times 100$ uniform grid in $x$ and $y$. Otherwise, a cubic scattered interpolation scheme is used.

\subsection{Sparse Computation of Velocity Gradients}

\begin{figure}[t!]
    \centering
    \includegraphics[width=1\textwidth]{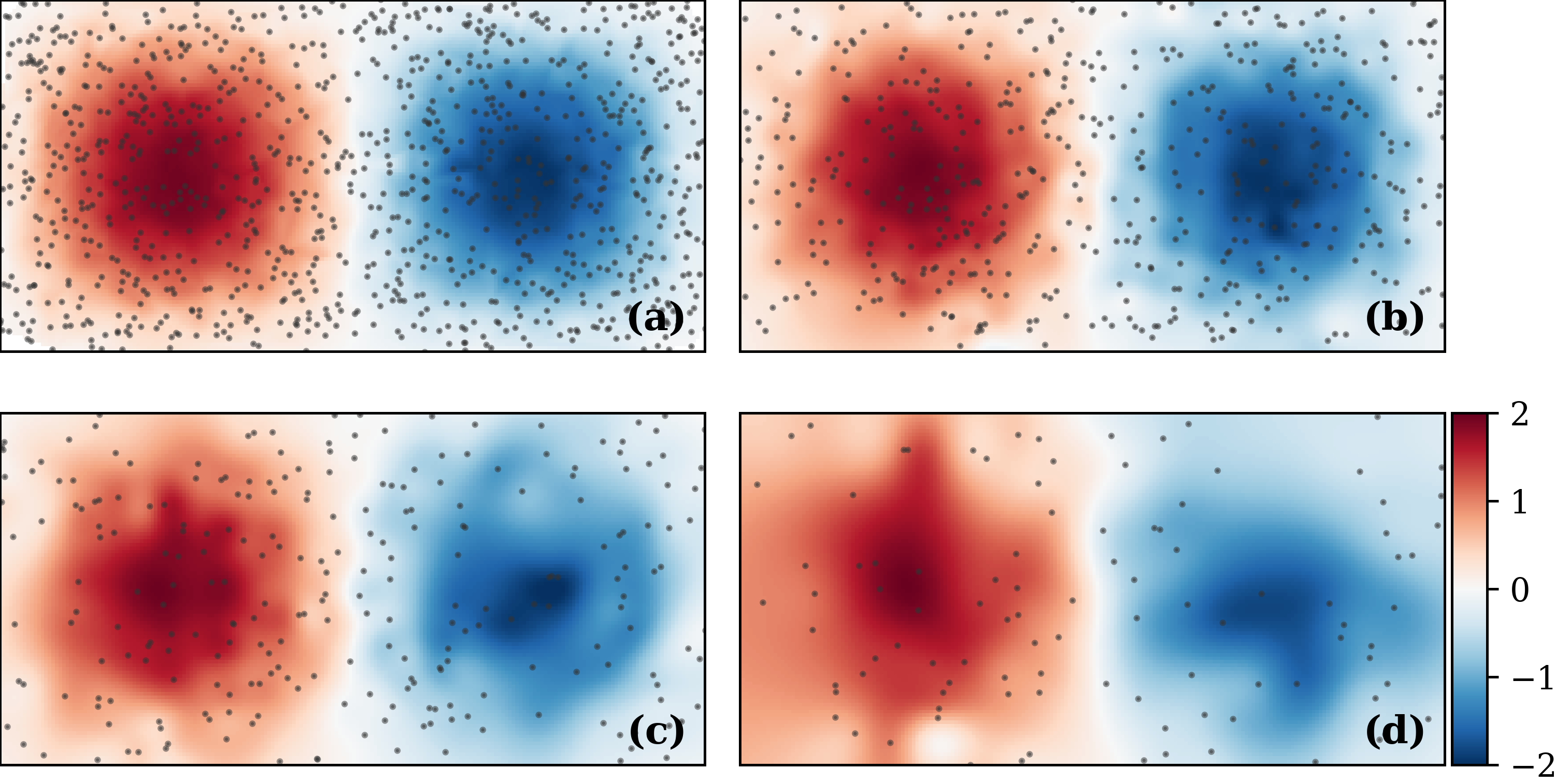}
    \caption{Vorticity computed on fields of sparsely distributed random particles using LGR.  The velocity gradient is computed via algorithm \ref{alg:lgr}, where $\F$ is computed over $t\in[0,\,0.1]$ using radial Gaussian weighting on the $k=15$ nearest neighbor particles.  Gray markers indicate the particle positions at $t_0=0$.  (a) 1000 particles.  (b) 500 particles.  (c) 250 particles.  (d) 100 particles.}
    \label{fig:Random_Vort}
\end{figure}

Lagrangian gradient regression is first demonstrated for its capacity to evaluate velocity gradients at an instance in time.  Using randomly distributed particle trajectories, algorithm \ref{alg:lgr} is implemented to compute $\nabla \mb{v}$, from which vorticity $\omega_z$ is computed.  For this demonstration, the regression time is over the domain $t_0 = 0$ to $t=0.1$ for all particle densities.  The results are presented in figure \ref{fig:Random_Vort} where figure \ref{fig:Random_Vort}(a) uses 1000 randomly distributed particles uses across the domain, figure \ref{fig:Random_Vort}(b) uses 500, figure \ref{fig:Random_Vort}(c) uses 250, and figure \ref{fig:Random_Vort}(d) uses 100.  The color mapping is chosen based on the analytically computed vorticity at $t_0=0$ in the double gyre flow, and is the same as the scheme used in figures \ref{fig:FtoL}(a, d, g).  The gray markers represent the particle positions at the evaluation time $t_0=0$.  

Figure \ref{fig:Random_Vort} demonstrates that vorticity---and more generally, the velocity gradient---can be accurately computed directly from Lagrangian trajectories recorded over short periods of time.  In figure \ref{fig:Random_Vort}(a), where the particle density is highest, the recorded vorticity matches very closely with the analytically computed field from figure \ref{fig:FtoL}(a) which was computed on a uniform grid consisting of twenty times as many particles.  In figures \ref{fig:Random_Vort}(b) and \ref{fig:Random_Vort}(c), which contain $1/40$\textsuperscript{th} and $1/80$\textsuperscript{th} as many particles as figure \ref{fig:FtoL}(a) respectively, the shape and magnitude of the vorticity field is consistent with the true values, though there are more errors due largely to interpolation effects.  Even in figure \ref{fig:Random_Vort}(d), where only 100 particles are used, the center of vorticity on either side of the vertical center line is localized to the correct location and the magnitude of the vorticity does not deviate far from the true values.

The ability to compute velocity gradients directly from sparse trajectories is advantageous in the context of sparse measurements.  Computing velocity gradients from tracer data can be difficult even when the recorded trajectories are abundant \cite{EtebariVlachos_DPIVDerivativeEstimation_2005, Beresh.Smith_ChallengesPIVTurbVisc_2018}.  When the data is sparse, the difficulty is compounded.  This challenge is typical when dealing with experimental flow-field measurements.  For example, in figures \ref{fig:Random_Vort}(a, b), the particle density is low, which limits the ability of PIV to achieve sufficient velocity field resolution for computing vorticity \cite{Raffel.Kompenhans_PIVbook_2018, KeaneAdrian_PIVAnalysisDensity_1992}.  If PTV were used, the local particle velocity fields would need to be interpolated onto a grid prior to differentiation which is expensive in terms of storage and computation time and can produce additional errors.  The ability to accurately compute velocity gradients directly from sparse particles may help improve various analyses that depend on gradient-based quantities by reducing computational effort and storage requirements and improving accuracy.  For a comparison of LGR with other velocity gradient computation pipelines common to experimental fluid mechanics, see \cite{Harms.McKeon_LGRISPIV_2023}.

\subsection{Sparse Computation of FTLE}
\begin{figure}[t!]
    \centering
    \includegraphics[width=1\textwidth]{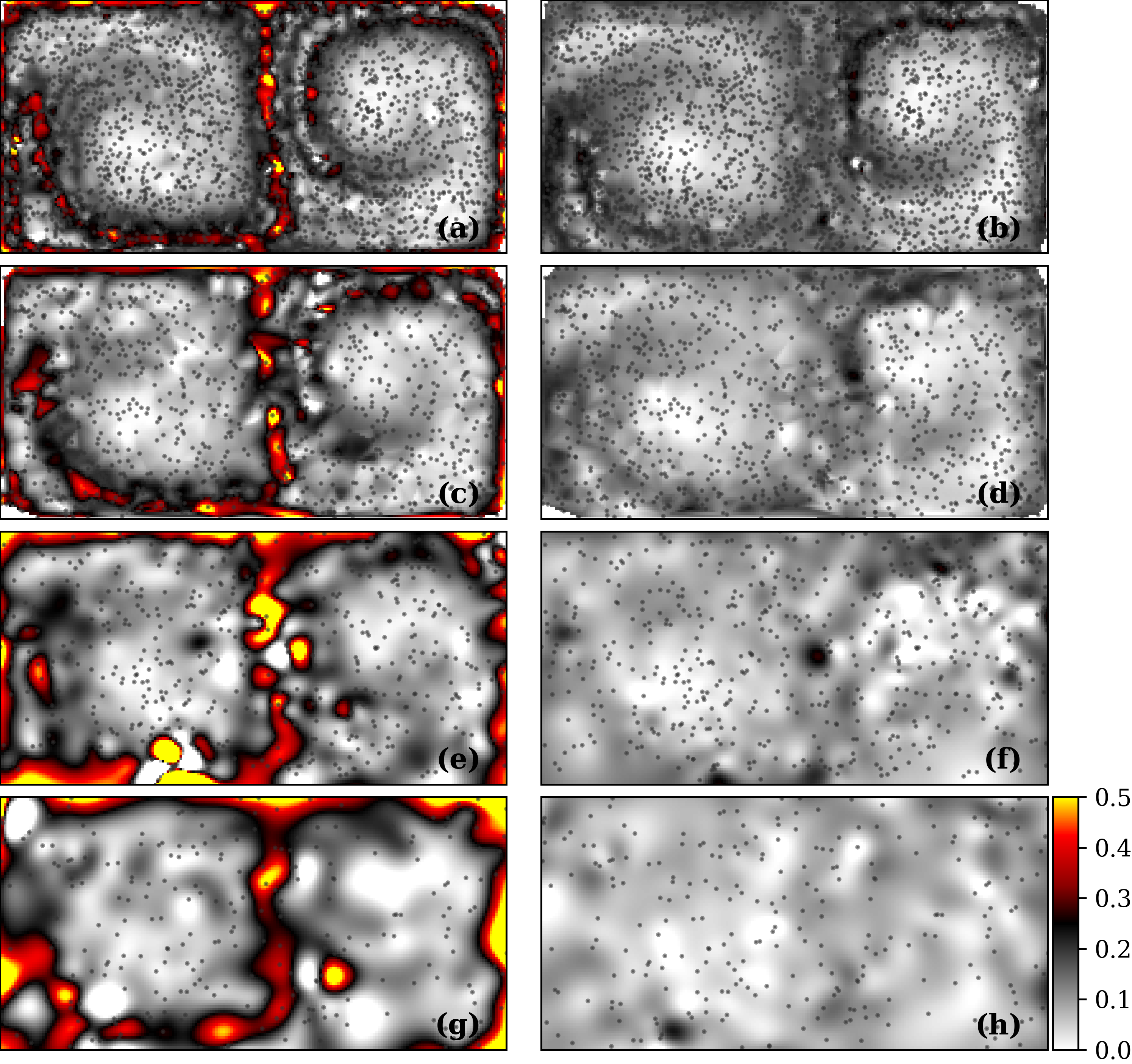}
    \caption{FTLE computed on fields of random particles using flow map composition with resampling (left column) and using only the initial and final positions of particles (right column). Particles are advected in $t\in[0,\,15]$ with resampling at $\Delta t =0.1$.  Jacobian computations use radial Gaussian weighting on the $k=15$ nearest neighbor particles.  Gray markers indicate the initial positions (evaluation locations) of the particles in the random field.  (a, b) 2500 particles.  (c, d) 1000 particles.  (e, f) 500 particles.  (g, h) 250 particles.}
    \label{fig:Random_FTLE}
\end{figure}

The second demonstration of the developed tools involves computing FTLE fields using algorithm \ref{alg:ftle} and comparing them with those generated using a traditional approach involving regression of the deformation between initial and final particle positions alone.  Both implementations identify the same $k=15$ nearest neighbors at time $t_0=0$ and perform analyses on the deformation up to $t=15$.  For the conventional approach, the Jacobian is computed using unweighted ordinary least-squares regression ($\mb{K} = \mb{I}_d$, $\gamma = 0$ in equation \ref{eq:regularizedkernelregression}). The results of the analysis are presented in figure \ref{fig:Random_FTLE}, where the left column (figure \ref{fig:Random_FTLE}(a, c, e, g)) displays the results using composition with replacement and the right column (figure \ref{fig:Random_FTLE}(b, d, f, h)) displays the traditional results.  Four particle densities are considered with increasing sparsity: 2500 particles in the domain(figure \ref{fig:Random_FTLE}(a, b)), 1000 particles in the domain (figure \ref{fig:Random_FTLE}(c, d)), 500 in the domain (figure \ref{fig:Random_FTLE}(e, f)), and 250 in the domain (figure \ref{fig:Random_FTLE}(g, h)).  This is in contrast with the density often used for FTLE computations on this flow, which can be as high as $10^7$ particles \cite{AllshousePeacock_LagrangianBasedMethods_2015}.  The values of the color mapping are chosen based on the true FTLE fields as seen in figure \ref{fig:LLR_Perf}, which readers should refer to as a reference example from a dense, structured field. The gray dots in the fields shown in figure \ref{fig:Random_FTLE} represent the initial particle positions at $t_0 = 0$ and are the scattered data points from which interpolation to the uniform grid is performed.  

The performance improvement of composition with resampling over typical Jacobian regression for sparse data is evident in figure \ref{fig:Random_FTLE}.  In each of the fields that compute the FTLE via LGR with composition (figures \ref{fig:Random_FTLE}(a, c, e, g)), some evidence of the principal FTLE ridges is present---even with as few as 250 particles embedded in the flow.  In contrast, only shadows of the ridges appear when the traditional approach is used, and those only when the field is more densely seeded (figures \ref{fig:Random_FTLE}(b, d)).  In figures \ref{fig:Random_FTLE}(f, h) there is no evidence of anything resembling a ridge in the FTLE field.  Additionally, the values observed in the FTLE fields computed via composition are commensurate with those in the true FTLE field for the observed flow (see figure \ref{fig:LLR_Perf}).  Therefore, the composition with resampling approach is a viable means of approximating FTLE fields from sparse data and qualitatively identifying hyperbolic LCS within them.

\subsection{Sparse Computation of Elliptic Metrics}

\begin{figure}[t!]
    \centering
    \includegraphics[width=1\textwidth]{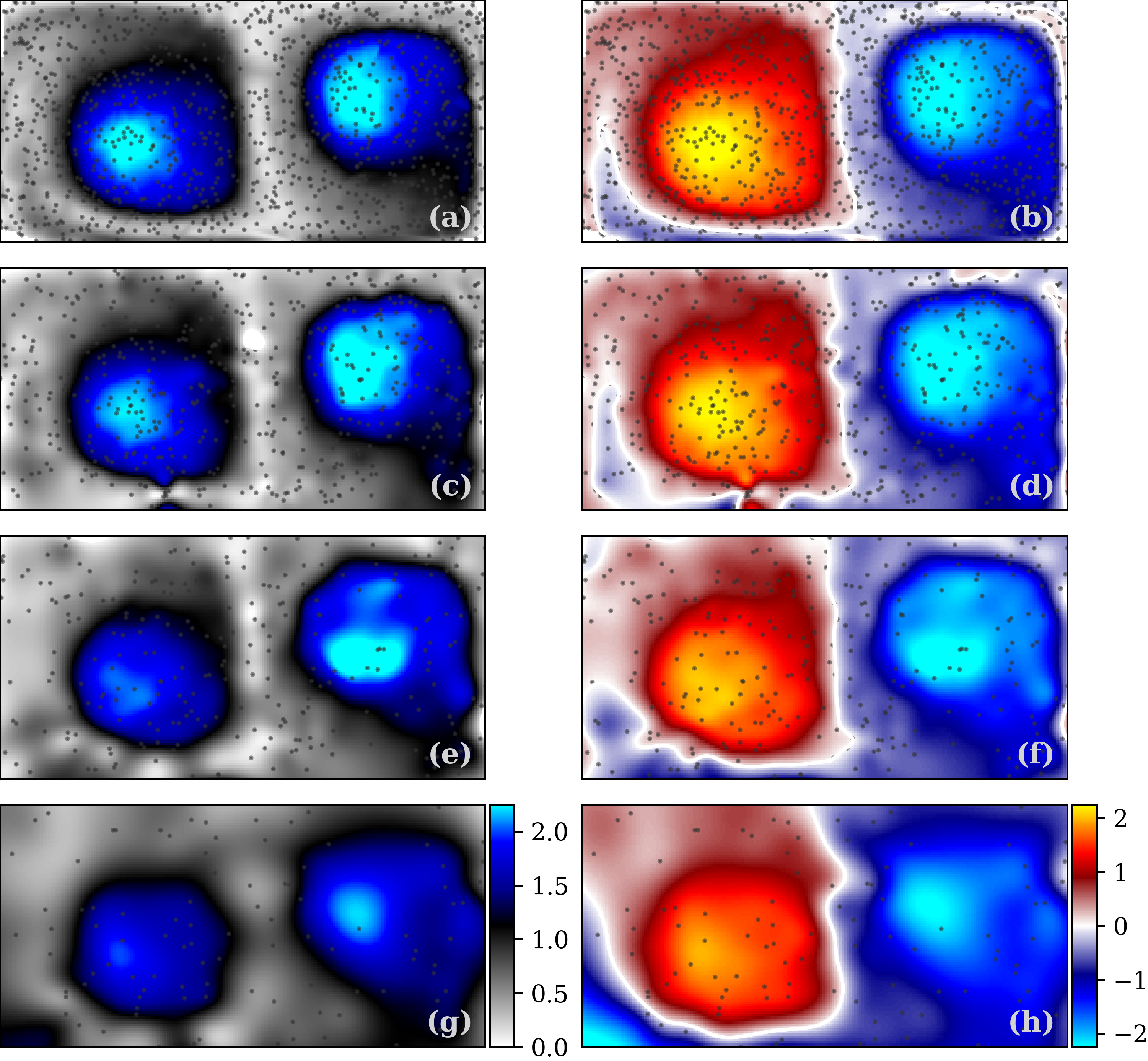}
    \caption{Lagrangian metrics for elliptic LCS identification computed from sparsely distributed fields of random particles.  Quantities are computed over $t\in[0,\,15]$ using regression of the $k=15$ nearest neighbor particles with radial Gaussian weighting. Nearest neighbor particles used in the regressions are replaced every $\Delta t = 0.1$.  Gray markers represent the particle positions at $t_0 = 0$.  First column: intrinsic rotation angle $\psi_{0}^{15}$. Second column: dynamic rotation angle $\varphi_{0}^{15}$. (a, b) 1000 particles.  (c, d) 500 particles. (e, f) 250 particles. (g, h) 100 particles.}
    \label{fig:Random_Ellip}
\end{figure}

The final demonstration computes elliptic LCS metrics from sparse trajectories using algorithm \ref{alg:lavd_dra}, the results of which are presented in figure \ref{fig:Random_Ellip}.  Specifically, the intrinsic rotation angle $\psi_{t_0}^t$ (scaled LAVD) and the dynamic rotation angle $\varphi_{t_0}^t$ are computed on the time domain $t\in [0,\,15]$.  In the column on the left (figures \ref{fig:Random_Ellip}(a, c, e, g)) results of $\psi_{t_0}^t$ are presented with increasing particle density, and in the right column (figures \ref{fig:Random_Ellip}(b, d, f, h)) results of $\varphi_{t_0}^t$ are displayed for the same data as the left column.  Computations are performed in figures \ref{fig:Random_Ellip}(a, b) using 1000 randomly seeded trajectories, in figures \ref{fig:Random_Ellip}(c, d) using 500, in figures \ref{fig:Random_Ellip}(e, f) using 250, and in figures \ref{fig:Random_Ellip}(g, h) using 100.   The color mapping represents the number of complete rotations observed by the measured quantity, and the maximum absolute value of the mapping is the same in both columns.  Additionally, the mapped values can be compared to those of figures \ref{fig:EllipMetrics}(a, c) which demonstrate $\psi_{t_0}^t$ and $\varphi_{t_0}^t$ respectively on dense trajectories.  As in figures \ref{fig:Random_Vort} and \ref{fig:Random_FTLE}, the gray dots represent the evaluation positions of the metrics at time $t_0=0$.

The results presented in figure \ref{fig:Random_Ellip} demonstrate that it is possible to accurately evaluate elliptic Lagrangian metrics from tracer data with no velocity information known \textit{a-priori}.  Moreover, elliptic LCS features are seen to be robust to sparsity.  In figures \ref{fig:Random_Ellip}(a, b) where 1000 random particles are used in the computations, both the IRA $\psi_{t_0}^t$ and DRA $\varphi_{t_0}^t$ closely resemble their dense counterparts in figure \ref{fig:EllipMetrics}.  As the tracer density decreases, the form of the structures in the field remains consistent so that, even with only 100 particles seeded in the domain---as in figures \ref{fig:Random_Ellip}(g, h)---the peaks and boundaries of the vortical regions of the flow are still readily observed.  

Another feature of interest in figure \ref{fig:EllipMetrics} is the zero-contour of the DRA $\varphi_{t_0}^t$.  As discussed in section \ref{sec:ellipLCS}, this contour represents the regions of the flow that experience net-zero rotation over the course of the flow.  Because this contour lies between regions of opposite rotation, it will always exist in DRA fields where multiple vortices are present and will continuously divide them.  Therefore, when dealing with sparse flows in practice, it may be convenient to use the zero-contour of the DRA as an approximation of hyperbolic LCS.  From figures \ref{fig:Random_Ellip}(b, d, f, h) it is evident that this contour (seen as a white line in between colored regions) is clearly identifiable even with few particles and that it approximately follows the FTLE ridges seen in figures \ref{fig:LLR_Perf} and \ref{fig:Random_FTLE}.  However, this contour does not exactly represent the FTLE ridges and is not objective.  Therefore it should only be considered as an approximate LCS.

\subsection{Structure Sensitivity and Robustness}

When comparing the LCS results from sparse trajectories to the those in previous sections where computations were performed on dense, structured data, it is apparent that the elliptic metrics and the velocity gradients are more robust to sparsity than the hyperbolic/parabolic LCS as revealed by the FTLE fields.  In figure \ref{fig:Random_FTLE}, for example, drawing precise ridges would be a challenging task to implement algorithmically for even the relatively dense field with 2000 random particles.  On the other hand, the salient features of the elliptic LCS (figure \ref{fig:Random_Ellip}) and velocity gradients (figure \ref{fig:Random_Vort}) are clearly visible with only 100 particles in the domain.  This discrepancy results from the topology of the computed structures and their sensitivity to interpolation errors; where elliptic LCS and velocity gradients are essentially measuring volumetric quantities, hyperbolic/parabolic LCS identify codimension-1 manifolds (material surface) of infinitessimal thickness that are more difficult to sense.  

\begin{figure}[t!]
    \centering
    \includegraphics[width=0.6\textwidth]{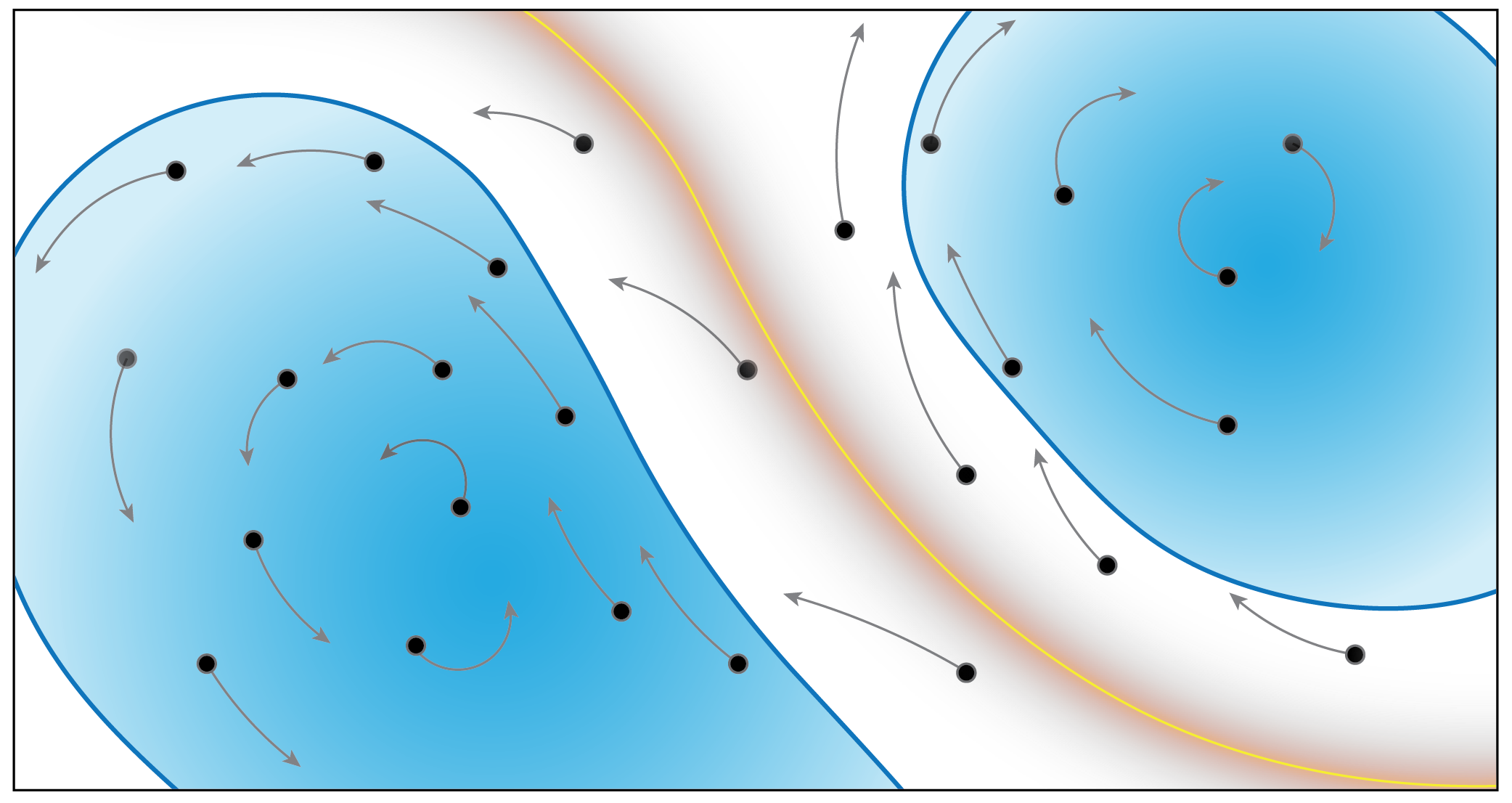}
    \caption{Illustration of geometric LCS emphasizing robustness to sparsity.  The orange curve represents a hyperbolic/parabolic LCS while the blue regions identify elliptic LCS.  Because many tracers may exist on the elliptic LCS, they are more easily sensed in a flow than the manifolds of attraction, repulsion, or shear. }
    \label{fig:Robustness}
\end{figure}

To illustrate this further, consider the schematic in figure \ref{fig:Robustness}.  Tracers and their trajectories are indicated alongside a repulsive hyperbolic LCS ridge (in orange) and two elliptic LCS (in blue) which hold tracers in Lagrangian vortices.  Because the hyperbolic ridge is infinitesimally thin, it is exceedingly unlikely that any tracers exist immediately on top of the feature.  Moreover, because the ridge is repulsive, the trajectories that begin near it diverge over time, allowing the nonlinear influence of the flow to skew results.   The elliptic LCS, however, entrap particles within their boundaries for the observed duration.  Since these particles exhibit similar rotational behavior and are spread over finite volume, the feature is more likely to be sensed by identification algorithms.  An implication of this is that elliptic LCS are more amenable to interpolation than hyperbolic LCS.  Because a larger region of tracers is affected by an elliptic LCS than a hyperbolic LCS, it is much simpler to define a field from it over sparse data.  These factors account for the improved robustness of elliptic LCS over hyperbolic and parabolic ones seen throughout this work, and supports the findings of other studies on LCS robustness such as that by Badza et al. (2023) \cite{Badza.Balasuriya_LCSSensitivity_2023}.

%% file: 6_SummaryAndConclusions.tex
\section{Conclusions \label{sec:conclusions}}

This work presents methods for identifying geometric Lagrangian coherent structures from sparse tracer data without any velocity field information.  The proposed methods are built upon the framework of Lagrangian gradient regression (LGR), which is introduced in section \ref{sec:timederivs} as a velocity- and derivative-free approach to computing velocity gradients from trajectories.  Where traditional approaches rely on finite-differences over a structured array of velocity information, LGR takes the data-scientific approach of regression.  A linear operator capturing particle deformation from one time instance to the next is regressed.  This operator is the flow map Jacobian over the time domain and, if the time step is small enough, the linear operator is connected to the velocity gradient by a simple algebraic equation.  Using regression rather than finite-differences enables a variety of algorithmic modifications to exploit different aspects of the data.  In this paper, kernel weighting and Tikhonov regularization are employed.  

To extend the framework of LGR to the study of geometric LCS, the composition property of the flow map is described and implemented in section \ref{sec:hyper_para}.  When using LGR to identify velocity gradients, the time step is necessarily small.  However, when computing LCS quantities such as the finite-time Lyapunov exponent (FTLE), arbitrarily long time domains must be considered.  Through the application of composition, LGR can be used to identify all intermediate flow map Jacobians, which can be multiplied together by application of the chain rule of calculus to achieve the long-time result.  With the complete flow map Jacobian computed, the typical machinery can be used to identify hyperobolic and parabolic LCS through, for example, FTLE fields.  Another result presented in section \ref{sec:hyper_para} involves computing long-time Jacobians along a single particle trajectory using known velocity gradients along its path.  This may be viewed as the inverse of the problem in LGR.  

LGR can also be used as a tool for computing Lagrangian-averaged vorticity deviation (LAVD) and dynamic rotation angle (DRA) directly from trajectory data which, to the authors' knowledge, has not yet been accomplished in the literature.  This is possible because LAVD and DRA are determined by the vorticity in a flow, which is computed in a Lagrangian sense (that is, field quantities are never computed) by LGR.  In section \ref{sec:ellipLCS}, it is shown how LAVD and DRA are computed as the trajectory average of the vorticity deviation magnitude and vorticity respectively, using LGR as the engine for computing the quantities at each instance in time.  Additionally, it was shown that the rate of change of both the DRA and the polar rotation angle (PRA) is the spin tensor $\mb{W}$, which implies that the dynamic rotation angle can be computed using either the vorticity or the sum of PRA computed at short intervals. 

These methods were seen to perform well on sparse trajectory data from the analytical double gyre flow model.  In section \ref{sec:randdata} it is shown that velocity gradients are accurately computed from sparse, random distributions via LGR; that incorporating composition with LGR enables FTLE fields to be computed with much greater accuracy than standard methods; and that LGR enables the direct computation of LAVD and DRA from sparse particle trajectories.  Moreover, it was observed that the volumetric quantities ($\nabla \mb{v}$ and elliptic metrics) were far more robust to sparsity than the infinitesimally thin hyperbolic structures in the FTLE ridges.

Computing the velocity gradients directly from trajectory data via LGR bucks the paradigm of requiring field information when identifying gradients.  This can be advantageous in multiple ways.  First, by remaining Lagrangian throughout the numerical process, LGR is far more storage efficient than field-based metrics.  Gradients need not ever be stored on dense grids, but can reside on the particle trajectories.  Another advantage is the introduction of the vast literature on regression to the task of gradient identification.  While only kernel-weighted least squares is used here, many other options for regression may be useful.   

In the context of computing LCS features, composition with LGR represents local-linearization in time rather than the conventional linearization in space.  It is this trade-off that allows for computation on sparse data.  By analyzing data that is sampled at a high resolution in time, it is not necessary to have trajectories that are sampled at a high resolution in space.  This has numerical advantages, as it allows for far fewer trajectories to be integrated and stored, but it is perhaps more advantageous when working with physical systems.  In laboratory and field experiments, sampling frequently in time is often more attainable than sampling at high resolution in space.  PTV, for example, requires high-frequency sampling in order to obtain the particle trajectories, and oceanic tracer data from drifters or ice floes are typically time-resolved with respect to the time-scales of the flow, but cannot be densely seeded due to practical concerns.  The Jacobian computation approach developed here is naturally suited to these applications and others like them.   

Since elliptic structures are seen to be more robust than hyperbolic and parabolic ones, it is expected that they will be of particular importance as autonomous technologies are built around flow-field information.  Not only do the tracers within these structures remain localized over long periods of time, they are more easily identified using fewer tracers.  This reduces the difficulties associated with sparsity and particle persistence in frame when performing computations.  Furthermore, when objectivity is not specifically required, the DRA may serve as a useful tool for approximating hyperbolic boundaries and judging the signed direction of material rotation.

There exist many avenues of development for this work.  First, and perhaps most apparent, is that these tools should be applied to a broad range of flows of varying complexity, sparsity, and dimension.  In this work the methods were demonstrated only on a simple analytical flow, but it is necessary that they be used on more complex problems.  Additionally, studies need to be undertaken characterizing robustness to measurement noise and the propagation of error.  It may also be beneficial to explore other variations of the regression problem (kernel regression, for example) for their viability in learning from Lagrangian data.  Regarding computation, the algorithms themselves could be further optimized, including parallelization.  Ultimately, however, the goal of this research is to use Lagrangian tracer information in the decision-making protocol of autonomous technologies.  Therefore, work is currently underway to assess the viability of the tools developed here for purposes of autonomy.  

\section{Acknowledgements}
The authors acknowledge funding from the US Army Research Office under grant number W911NF-17-1-0306 and from the US Office of Naval Research under grant N0014-17-1-3022.

%% file: A_Appendices.tex
\appendix

\section{Proof of Equation \ref{eq:linearProcessJacobian}}

In section \ref{sec:timederivs} it is taken for granted that the velocity gradient serves as the evolution function of the linear process defining the flow map Jacobian.  Indeed, this is understood in the continuum mechanics literature \cite{Gurtin.Anand_ContinuumMechanics_2010, TruesdellRajagopal_IntroToFluids_2000, Haller_DynamicPolarDecompostion_2016}, but the proof is worth elaborating.  This appendix provides a form of the derivation of equation \ref{eq:linearProcessJacobian}.  

Intrinsic to the definition of a flow map (equation \ref{eq:flowmap}) is a reference frame against which all tracer deformations are measured.  The reference frame is arbitrary and is typically taken to be the initial positions of the particles being examined.  There are instances, however, when it is useful to understand tracer deformation with respect to another position, as with the intermediate flow maps used in compositions.  It is for such situations the relative flow map is useful :
\begin{equation}
    \mb{F}_{t}^{\tau}(\mb{x}(t)) = \mb{F}_{t}^{\tau}(\mb{F}_{t_0}^{t}(\mb{x}_0)) = \mb{x}(\tau; t, \mb{F}_{t_0}^{t}(\mb{x}_0)).
    \label{eq:relflowmap}
\end{equation}
From here the relative flow map Jacobian is defined (this is the relative deformation gradient in the continuum mechanics language \cite{Gurtin.Anand_ContinuumMechanics_2010})
\begin{equation}
    D\mb{F}_{t}^{\tau}(\mb{x}(t)) = \nabla_{\mb{x}(t)} \mb{F}_{t}^{\tau}(\mb{x}(t)),
\end{equation}
where $\nabla_{\mb{x}(t)}$ represents the gradient with respect to the tracer position at time $t$.  

When $\Delta t = \tau-t \ll \mathcal{T}$, where $\mathcal{T}$ is the fast time scale of the flow, it is possible to approximate the temporal derivative of the the flow map Jacobian.  Thus, under these conditions,
\begin{equation}
    \frac{D\mb{F}_{t_0}^{\tau}(\mb{x}_{0})-D\mb{F}_{t_0}^{t}(\mb{x}_{0})}{\Delta t} = \dfrac{d}{d t} D\mb{F}_{t_0}^{t}(\mb{x}_{0}).
\end{equation}
Then, by applying the process property,
\begin{equation}   
     \dfrac{d}{d t}D\mb{F}_{t_0}^{t}(\mb{x}_{0}) = \left(\dfrac{d} {d \tau} \left.D\mb{F}_{t}^{\tau}(\mb{x}_t)\right\rvert_{\tau=t}\right) D\mb{F}_{t_0}^{t}(\mb{x}_{0}),
     \label{eq:diffRelFMJ1}
\end{equation}
where
\begin{equation}
    \dfrac{d} {d \tau} \left.D\mb{F}_{t}^{\tau}(\mb{x}_t)\right\rvert_{\tau=t} \approx \frac{D\mb{F}_{t}^{\tau}(\mb{x}_{t})-\mb{I}_{d}}{\Delta t}
    \label{eq:diffRelFMJ2}
\end{equation}
is the rate of change of the relative flow map Jacobian.  

A well-established result in continuum mechanics relates the temporal derivative of the flow map Jacobian to the velocity gradient \cite{Gurtin.Anand_ContinuumMechanics_2010}.  To see this, consider the rate of change of the Jacobian.
\begin{equation}
    \dfrac{d}{d \tau} \left.D\mb{F}_{t}^{\tau}(\mb{x}(t))\right \rvert_{\tau=t} = 
    \dfrac{d}{d \tau} \nabla_{\mb{x}(t)} \left.\mb{F}_{t}^{\tau}(\mb{x}(t))\right \rvert_{\tau=t} = 
    \nabla_{\mb{x}(t)} \dfrac{d}{d \tau} \left.\mb{F}_{t}^{\tau}(\mb{x}(t))\right \rvert_{\tau=t} = \nabla_{\mb{x}(t)} \mb{v}(\mb{x}(t),t),
    \label{eq:diffRelFMJ3}
\end{equation}
where the last step was achieved by applying equation \ref{eq:velocity}. Inserting the result into equation \ref{eq:diffRelFMJ1} yields the evolution equation of a linear process for the flow map Jacobian
\begin{align}
    \dfrac{d}{d t} \F(\mb{x}_0) &= \nabla \mb{v}(\mb{x}(t),t) \F(\mb{x}_0),
\end{align}
where $D\mb{F}_{t_0}^{t_0} = \mb{I}_{d}$.

\section{Implications of Objectivity \label{ap:objectivity}}

A hallmark of the theory of Lagrangian coherent structures is the property of objectivity, which ensures that the computed quantities remain consistent regardless of the motion of the observer.  While it has been briefly touched on in the body of the paper, objectivity has not yet been thoroughly discussed in this work, but rather assumed.  In this appendix, various discussions of objectivity from relevant works are synthesized to provide an overview of the objectivity of Lagrangian quantities.  With the exception of the discussion regarding the objectivity of the composition operation, all of the results can be found throughout the literature.  The purpose of this appendix is to summarize the objectivity of some relevant Lagrangian quantities and operations in one location so that the reader does not need to sift the literature for them.  Readers interested in greater depth should refer to the helpful chapter on objectivity from Haller's recent textbook on LCS \cite{Haller_LCStextbook_2023} and to many articles touching the subject \cite{Gurtin.Anand_ContinuumMechanics_2010, TruesdellRajagopal_IntroToFluids_2000, Haller_ObjectiveDefVortex_2005, Haller_DynamicPolarDecompostion_2016, FarazmandHaller_PolarRotationAngle_2016}, among others.  

We say that a quantity is objective if it exhibits invariance under Euclidean transformations of the form 
\begin{equation}
    \tilde{\mb{x}}(t) = \mb{Q}(t)\mb{x}(t) + \mb{p}(t),
    \label{eq:changeofframe}
\end{equation}
where $\mb{Q}(t)$ is a proper orthogonal rotation tensor and $\mb{p}(t)$ is a translation.  The objectivity of a scalar, vector, or tensor quantity is examined by considering the influence of changes of frame in the form of equation \ref{eq:changeofframe} on the resulting value.  To aid the discussion, some definitions from continuum mechanics are useful \cite{Gurtin.Anand_ContinuumMechanics_2010}.  
\begin{definition}[Frame-Indifference]
    A scalar field $g$ is frame-indifferent if it is unchanged by frame rotation and translation
    \begin{equation}
        \tilde{g} = g,
    \end{equation}
    where $\tilde{\cdot}$ represents the transformed quantity.  Moreover, a vector field $\mb{g}$ is frame-indifferent if it simply rotates with the frame rotation
    \begin{equation}
        \tilde{\mb{g}} = \mb{Qg},
    \end{equation} 
    and a tensor field $\mb{G}$ is frame-indifferent if, given frame-indifferent vector fields $\mb{g}$ and $\mb{h}$, 
    \begin{equation}
        \mb{h = Gg} \implies \tilde{\mb{h}}=\tilde{\mb{G}}\tilde{\mb{g}}.
    \end{equation}
    Then, the transformation law for a frame-indifferent tensor field is
    \begin{equation}
        \tilde{\mb{G}} = \mb{QGQ}^\top.
    \end{equation}
\end{definition}
In other words, a frame indifferent vector is one such that its magnitude remains unchanged by the change of frame, and a frame indifferent tensor is one that maps indifferent vectors into indifferent vectors \cite{TruesdellRajagopal_IntroToFluids_2000}.  

The concept of invariance is related to that of frame-indifference.  
\begin{definition}[Invariance]
    A vector field or tensor field is invariant if it remains unchanged under transformations of the form presented in equation \ref{eq:changeofframe}.  Therefore, any invariant quantity $\mb{H}$ obeys the transformation law
    \begin{equation}
        \tilde{\mb{H}}= \mb{H}.
    \end{equation}
    Furthermore, any scalar valued function of a tensor $f(\mb{G})$ is invariant if and only if 
    \begin{equation}
        f(\mb{QGQ}^\top) = f(\mb{G})\quad \forall |\det \mb{Q}| = 1.
    \end{equation}
    A quantity that is invariant is objective.  
\end{definition}

Using these definitions, the objectivity of $\F$, $\C$, and $\ftle$ is readily assessed \cite{Gurtin.Anand_ContinuumMechanics_2010, Shadden.Marsden_FTLEProperties_2005}.  Using the definition of the flow map from equation \ref{eq:flowmap}, the transformed flow map is given
\begin{align}
\begin{split}
    \tilde{\mb{x}}(t) = \mb{Q}(t)\mb{x}(t) + \mb{p}(t) &= \mb{Q}(t)\mb{F}_{t_0}^{t}\left(\mb{x}_0\right) + \mb{p}(t).
\end{split}
\end{align}
Computing the gradient with respect to $\mb{x}_0$ yields
\begin{equation}
    \Ft = \mb{Q}(t)\F.
    \label{eq:transF}
\end{equation}
Hence, $\F$ is neither objective nor frame-indifferent.  This result can be used to see that $\C$ is objective:
\begin{equation}
    \Ct = \left(\mb{Q}(t)\F\right)^\top \mb{Q}(t)\F = (\F)^\top \mb{Q}^\top(t)\mb{Q}(t)\F = (\F)^\top\F = \C.
\end{equation}
Then, since $\C$ is objective, $\ftle$ is also objective.

\subsection{Objectivity of the Composition Operation}

Computing the FTLE using initial and final times alone is objective because the analysis is based on the right Cauchy-Green tensor which, as has been shown, is invariant.  However, computing the FTLE via composition according to equation \ref{eq:compeq} depends on many instances of the relative deformation gradient, which is not objective.  For a single instance of composition, 
\begin{align}
    \F = D\mb{F}_s^t D\mb{F}_{t_0}^s, 
\end{align}
from which an expression for the relative Jacobian is obtained
\begin{align}
    D\mb{F}_s^t = \F \left(D\mb{F}_{t_0}^s\right)^{-1}. 
\end{align}
Under a change of reference, this is expressed as
\begin{align}
\begin{split}
    D\tilde{\mb{F}}_s^t &= \Ft \left(D\tilde{\mb{F}}_{t_0}^s\right)^{-1} \\
    &= \mb{Q}(t)\F\left(\mb{Q}(s)D\mb{F}_{t_0}^s\right)^{-1} \\
    &= \mb{Q}(t)\F D\mb{F}_s^{t_0}\mb{Q}^\top(s)\\
    &= \mb{Q}(t)D\mb{F}_s^t\mb{Q}^\top(s). 
\end{split}
\end{align}
Therefore, in general,
\begin{equation}
    D\tilde{\mb{F}}_{t_i}^{t_{i+1}} = \mb{Q}(t_{i+1})\mb{F}_{t_i}^{t_{i+1}}\mb{Q}^\top(t_i).
\end{equation}
Applying this to equation \ref{eq:compeq}, an expression for the composite flow map Jacobian under a change of reference is
\begin{align}
\begin{split}
    D\tilde{\mb{F}}_{t_0}^{t_n} (\mb{x}_0) &= \prod_{i=0}^{n-1} D\tilde{\mb{F}}_{t_{i}}^{t_{i+1}} \left( \mb{x}(t_{i}) \right) \\ 
    &=\prod_{i=0}^{n-1}\mb{Q}(t_{i+1})D\mb{F}_{t_{i}}^{t_{i+1}} \left( \mb{x}(t_{i}) \right) \mb{Q}^\top(t_{i}), \\
    \begin{split}\; = \mb{Q}(t_{n}) D\mb{F}_{t_{n-1}}^{t_{n}} (\mb{x}(t_{n-1}))\mb{Q}^\top(t_{n-1}) \mb{Q}^\top(t_{n-1}) D\mb{F}_{t_{n-2}}^{t_{n-1}} (\mb{x}(t_{n-2}))\mb{Q}^\top(t_{n-2}) \cdots \\ \mb{Q}(t_{1})^\top \mb{Q}(t_{1}) D\mb{F}_{t_{1}}^{t_{0}} (\mb{x}(t_{0}))\mb{Q}^\top(t_{0}) \end{split}\\
    &= \mb{Q}(t_{n})\prod_{i=0}^{n-1} D\mb{F}_{t_{i}}^{t_{i+1}} \left( \mb{x}(t_{i}) \right) \\
    &= \mb{Q}(t_{n})D\mb{F}_{t_0}^{t_n} (\mb{x}_0),
\end{split}
\end{align}
which is identical to $D\tilde{\mb{F}}_{t_0}^{t_n}$ computed from the only the first and last time instances.  Thus, the operation of composition does not change the objectivity of the flow map Jacobian or associated quantities.  

\subsection{Objectivity of the polar decomposition}
As discussed in section \ref{sec:ellipLCS}, the polar decomposition separates the flow map Jacobian into a proper orthogonal rotation tensor $\mb{R}_{t_0}^t$ and a symmetric positive definite right stretch tensor $\mb{U}_{t_0}^t$ or left stretch tensor $\mb{V}_{t_0}^t$ such that 
\begin{equation*}
    \F = \mb{R}_{t_0}^t \mb{U}_{t_0}^t = \mb{V}_{t_0}^t \mb{R}_{t_0}^t.
\end{equation*}
A property of the right stretch tensor $\mb{U}_{t_0}^t$ is that $(\mb{U}_{t_0}^t)^2=\mb{C}$.  As a result, the right stretch tensor is objective.  Then, if we consider the transformation of the flow map Jacobian, 
\begin{equation}
    \Ft = \mb{Q}\F = \mb{Q}\mb{R}_{t_0}^t\mb{U}_{t_0}^t,
\end{equation}
and $\tilde{\mb{R}}_{t_0}^t = \mb{Q}\mb{R}_{t_0}^t$.  Moreover, it follows that
\begin{equation}
    \tilde{\mb{V}}_{t_0}^t = \mb{Q}\F (\mb{R}_{t_0}^t)^{-1}\mb{Q}^\top = \mb{Q}\mb{V}_{t_0}^t\mb{Q}^\top.
\end{equation}
Therefore, the right stretch tensor is objective, the left stretch tensor is frame invariant, but not objective, and the rotation tensor is neither.  

The objectivity of the dynamic polar decomposition is also important to consider, and has been thoroughly documented by Haller (2016) \cite{Haller_DynamicPolarDecompostion_2016}.  Thus, those results will not be repeated here.  

\subsection{Objectivity of $\gradv$, $\mb{W}$, and $\mb{D}$}

The objectivity of the velocity gradient and its spin and dilatation (often called stretch or stretching) components begins by considering the objectivity of the rate of change of the flow map Jacobian \cite{TruesdellRajagopal_IntroToFluids_2000}.  Using the chain rule and abbreviating $\mb{Q}(t)$ as $\mb{Q}$, 
\begin{equation}
    \frac{d}{dt}\Ft = \mb{Q}\left(\frac{d}{dt}\F\right) + \left(\frac{d}{dt}\mb{Q}\right)\F.
\end{equation}
Now, by equation \ref{eq:linearProcessJacobian}, $\frac{d}{dt}\F = \gradv (\F)$.  Inserting this into the above and applying equation \ref{eq:transF} yields
\begin{equation}
    \tilde{\nabla \mb{v}} \left(\Ft\right) = \mb{Q} \nabla \mb{v} \left(\mb{Q}^\top \Ft\right) + \frac{d}{dt}\mb{Q}\left(\mb{Q}^\top \Ft\right).
\end{equation}
Since $\Ft$ is invertible, it can be removed from the equation
\begin{equation}
    \nabla \tilde{\mb{v}}= \mb{Q} (\nabla \mb{v}) \mb{Q}^\top + \mb{\Omega},
    \label{eq:gradv_transform}
\end{equation}
where the the frame spin is defined as
\begin{equation}
    \mb{\Omega} = \frac{d}{dt}\mb{Q}\mb{Q}^\top,
\end{equation}
and is a skew tensor defining the rate of rotation of the observer.  Here, equation \ref{eq:gradv_transform} represents the transformation equation for the velocity gradient.  Therefore, the velocity gradient is neither objective nor frame-indifferent.  

Splitting this into its components, 
\begin{equation}
    \tilde{\mb{D}} + \tilde{\mb{W}} = \mb{Q} (\mb{D} + \mb{W}) \mb{Q}^\top + \mb{\Omega}
\end{equation}
Since both $\mb{W}$ and $\mb{\Omega}$ are skew, and $\mb{D}$ is symmetric, the transformation equations for $\mb{D}$ and $\mb{W}$ are given as 
\begin{align}
    \tilde{\mb{D}} &= \mb{Q} \mb{D} \mb{Q}^\top,\\
    \tilde{\mb{W}} &= \mb{Q} \mb{W} \mb{Q}^\top + \mb{\Omega}.
\end{align}
Therefore, the dilatation tensor $\mb{D}$ is frame-indifferent but not objective, and the spin tensor $\mb{W}$ is neither frame-indifferent nor objective.  As a consequence, the principal strain as defined in equation \ref{eq:principal_stretch} is objective and the vorticity is not.  

\subsection{Objectivity of metrics for material rotation}
Using the transformation laws for the dilatation and spin tensors allows for transformation laws of scalar rotation metrics to be derived.  As discussed with equation \ref{eq:vorticity}, the vorticity at the location of a particle along its trajectory may be computed according to 
\begin{equation*}
    \mb{We} = -\frac{1}{2}\mb{\omega}\times \mb{e}, \quad \forall \mb{e}\in \R^d.
\end{equation*}
Applying the transformation, one obtains
\begin{align}
    \tilde{\mb{W}}\mb{e} &= (\mb{Q} \mb{W} \mb{Q}^\top + \mb{\Omega})\mb{e} \nonumber \\ 
    -\frac{1}{2}\tilde{\mb{\omega}}\times \mb{e} &= \mb{Q} (-\frac{1}{2}{\mb{\omega}}\times \mb{e}) - \frac{1}{2}\dot{\mb{q}}\times \mb{e} \nonumber \\
    \tilde{\mb{\omega}}\times \mb{e} &= \mb{Q} {\mb{\omega}}\times \mb{e} + \dot{\mb{q}}\times \mb{e} \nonumber \\
    \tilde{\mb{\omega}} &= \mb{Q} \mb{\omega} + \dot{\mb{q}}.
\end{align}
where $\dot{\mb{q}}$ is a vector representation of rate of frame rotation.  Furthermore, the transformation law for $Q$ from the $Q$-criterion (equation \ref{eq:Q-crit}) can be assessed.  
\begin{align}
    2\tilde{Q} &= \tilde{\norm{\mb{W}}}_F^2 - \tilde{\norm{\mb{D}}}_F^2 \nonumber \\
    &= \tr({\tilde{\mb{W}}^\top \tilde{\mb{W}}}) - \tr({\tilde{\mb{D}}^\top \tilde{\mb{D}}}) \nonumber \\
    &= \tr((\mb{QWQ}^\top + \mb{\Omega})^\top(\mb{QWQ}^\top + \mb{\Omega})) - \tr((\mb{QDQ}^\top)^\top(\mb{QDQ}^\top)) \nonumber \\
    &= \tr(\mb{QW}^\top \mb{WQ}^\top) - \tr(\mb{QD}^\top \mb{DQ}^\top) + 2\tr(\mb{QWQ}^\top \mb{\Omega}) + \tr(\mb{\Omega}^\top \mb{\Omega}) \nonumber \\
    &= 2Q + 2\tr(\mb{QWQ}^\top \frac{d}{dt}\mb{Q}\mb{Q}^\top) + \norm{\mb{\Omega}}_F^2 \nonumber \\
    &= 2Q - 2\tr(\mb{W}^\top \mb{\Omega}) + \norm{\mb{\Omega}}_F^2 \nonumber \\
    \tilde{Q} &= Q - \langle \mb{W}, \mb{\Omega} \rangle_F + \norm{\mb{\Omega}}_F^2,
\end{align}
where $\langle \cdot \rangle_F$ represents the Frobenius inner product. Here, the term $\norm{\mb{\Omega}}_F^2$ represents the magnitude of the frame rotation irrespective of the flow and the term $\langle \mb{W}, \mb{\Omega} \rangle_F$ represents the magnitude of the relative rotation between the flow and the observer.  Because the transformation laws of both vorticity and $Q$-criterion indicate that the observed value varies as a function of frame rotation, they are not objective.  This is intuitive, as one would expect that the observed rotation of a flow would appear different if observed from a rotating vantage.  

The objectivity of the LAVD (also the IRA) and the DRA are thoroughly discussed in Haller \cite{Haller_DynamicPolarDecompostion_2016} and Haller et. al. \cite{Haller.Huhn_LAVD_2016} where they are developed.  The LAVD and the IRA $\psi_{t_0}^t$ are both objective and dynamically consistent in two and three dimensions.  On the other hand, the DRA $\varphi_{t_0}^t$ is not objective.  In \cite{FarazmandHaller_PolarRotationAngle_2016}, Farazmand and Haller remark that closed level sets of the polar rotation angle are objective, though it is not in general.  For more information about the usage and objectivity of these quantities, refer to the sources listed above.

\begin{table}
    \centering
    \small
    \renewcommand{\arraystretch}{1.2} 
    \caption{Summary of objectivity and transformation laws for quantities of importance in the study of geometric LCS}
    \begin{tabular}{x{4cm}|x{1.5cm}x{2cm}x{2cm}|x{4.5cm}}
    \hline Quantity  & Objective & Frame-Indifferent & Dynamically Consistent & Transformation Law  \\  \hline
    Flow map Jacobian                   & \xmark & \xmark & N/A    & $\Ft = \mb{Q} \F$\\
    Flow map Jacobian (by composition)  & \xmark & \xmark & N/A    & $\Ft = \mb{Q}\prod_{i=0}^{n-1} D{\mb{F}}_{t_{i}}^{t_{i+1}} \left( \mb{x}(t_{i}) \right)$ \\
    Cauchy-Green tensor                 & \cmark & \cmark & N/A    & $\tilde{\mb{C}}_{t_0}^t = \C$\\
    Rotation tensor                     & \xmark & \xmark & N/A    & $\tilde{\mb{R}}_{t_0}^t = \mb{QR}_{t_0}^t$\\
    Right stretch tensor                & \cmark & \cmark & N/A    & $\tilde{\mb{U}}_{t_0}^t = \mb{U}_{t_0}^t$\\
    Left stretch tensor                 & \xmark & \cmark & N/A     & $\tilde{\mb{V}}_{t_0}^t = \mb{QV}_{t_0}^t \mb{Q}^\top$\\
    Velocity gradient                   & \xmark & \xmark & N/A    & $\nabla\tilde{\mb{v}} = \mb{Q} \nabla\mb{v} \mb{Q}^\top + \mb{\Omega}$\\
    Spin tensor                         & \xmark & \xmark & N/A    & $\tilde{\mb{W}} = \mb{Q} \mb{W} \mb{Q}^\top + \mb{\Omega}$\\
    Dilatation tensor                   & \xmark & \cmark & N/A    & $\tilde{\mb{D}} = \mb{Q} \mb{D} \mb{Q}^\top$\\
    FTLE                                & \cmark & \cmark & N/A    & $\tilde{\sigma}_{t_0}^t = \sigma_{t_0}^t$\\
    Vorticity                           & \xmark & \xmark & N/A    & $\tilde{\mb{\omega}} = \mb{Q} \mb{\omega} + \dot{\mb{q}}$ \\
    $Q$-Criterion                       & \xmark & \xmark & N/A    & $\tilde{Q} = Q - \langle \mb{W}, \mb{\Omega} \rangle_F + \norm{\mb{\Omega}}_F^2$\\
    Intrinsic rotation angle (LAVD)     & \cmark & \cmark & \cmark & $\tilde{\psi}_{t_0}^t = \psi_{t_0}^t$\\
    Polar rotation angle                & \xmark & \xmark & \xmark & --- \\
    Dynamic rotation angle              & \xmark & \xmark & \cmark & --- \\
    \end{tabular}
    \label{tab:my_label}
\end{table}

\clearpage